\documentclass[aps,prapplied,reprint,superscriptaddress,longbibliography]{revtex4-2}

\usepackage{graphicx}
\usepackage{amsmath,amssymb,amsfonts}
\usepackage{bm}
\usepackage{hyperref}
\usepackage{booktabs}
\usepackage{algorithm}
\usepackage{algpseudocode}
\usepackage{subcaption}
\usepackage{tikz}
\usepackage{multirow}
\usepackage{capt-of}
\usepackage{enumitem}
\usepackage{float}
\usetikzlibrary{matrix,shapes.geometric,decorations.pathmorphing,decorations.pathreplacing,patterns,decorations.markings}

\begin{document}

\title{Multiuser QKD using quotient graph states derived from continuous-variable dual-rail cluster states}

\author{Akash Nag Oruganti}
\email{akash.nag.10@gmail.com}
\affiliation{Department of Optics, Palack\'y University, Olomouc 77146, Czech Republic}

\date{\today}

\begin{abstract}
Multipartite entangled states are essential for multi-user quantum cryptography. While large-scale continuous-variable (CV) cluster states, particularly the dual-rail cluster state, are well-studied in measurement-based quantum computation, their cryptographic potential remains underexplored. Here, we propose a three-user conference key protocol using a CV dual-rail cluster state. By applying a node-coloring scheme to the infinite dual-rail graph, we create a six-mode pure graph state ideal for cryptographic tasks. Our results demonstrate near-GHZ performance for quantum conference key agreement (QCKA). Crucially, our protocol uniquely enables bipartite keys post-QCKA, which GHZ states cannot provide. It also surpasses two-mode squeezed vacuum states in generating bipartite keys within downstream access networks. Considering finite-size effects and impure squeezed states, our scheme remains robust despite experimental imperfections. We also introduce an enhanced method to more accurately estimate bipartite key generation capacity in quantum networks, paving the way for practical multi-user quantum cryptography.
\end{abstract}

\maketitle

\maketitle

\section{Introduction}
Multipartite entangled states are essential resources for various multi-user quantum cryptographic tasks, such as quantum voting \cite{votingGHZ, votingCV, votingdv}, quantum secret sharing \cite{QSScluster, QSScvgraph, QSSentanglement}, multipartite quantum direct communication \cite{QSDCreview, QDCwithGHZ}, and quantum conference key agreement (QCKA) \cite{QCKAreview, CV_CKA_exp}, among others. These tasks leverage the unique properties of entanglement to enable secure and efficient communication protocols in quantum networks.

In the realm of CV quantum systems, the development and application of multipartite entangled states have predominantly focused on measurement-based quantum computation, particularly using cluster states \cite{Menicucci2006,Menicucci2014,Larsen2021,qc1,qc2}. Despite significant technological advances in the generation of large-scale CV cluster states, especially the dual-rail cluster state, these resources have not been thoroughly explored for quantum cryptographic applications. This gap is notable given the experimental maturity achieved in creating such states through techniques like time \cite{dual-time-1, dual-time-2, timedualrail,dual-time4} and frequency domain multiplexing \cite{freq-1, freq-2, freq-3, freq-4}.

In discrete-variable (DV) systems, Greenberger-Horne-Zeilinger (GHZ) and W states have been identified as ideal resources for QCKA \cite{QCKAreview, DV_GHZ}. In CV systems, GHZ and W states are effectively equivalent, both represented by fully symmetric multimode squeezed states \cite{threemodecv}. However, it remains uncertain whether these CV equivalents, produced with finite squeezing, are the most suitable multipartite states for QCKA. Unlike DV systems, CV systems offer unique advantages, such as the ability to utilize conditional data post-QCKA to derive additional bipartite keys. This multifaceted use of data enhances the efficiency and security of quantum communication protocols, a feature we explore extensively in this work.

In this paper, we address this gap by introducing a novel protocol for generating a three-user conference key using a CV dual-rail cluster state. To enable its application in QCKA, we introduce the concept of a quotient graph state, employing a node coloring scheme on the infinite dual-rail graph. By forming the quotient graph through vertex identification based on this coloring, we obtain a six-vertex graph representing a pure six-mode graph state. This concept is new and provides a method for state engineering tailored for quantum communication applications, leveraging the properties of the dual-rail cluster state.

The underlying concept of a "quotient graph" is a well-established tool in graph theory \cite{godsil2001algebraic} used to simplify complex graphs by partitioning vertices. This technique has found applications in quantum mechanics for studying symmetries in quantum graphs \cite{band2017quotients,Mutlu2021,PhysRevA.75.062332}. Our work extends this principle to a new domain: we apply the quotienting procedure not merely for analysis, but as a constructive method to engineer a target multipartite entangled state from a larger, periodic resource state. This approach retains the essential entanglement and connectivity properties required for cryptographic protocols.

We investigate three schemes for generating QCKA in a three-user network:

Direct Reconciliation: The dealer (the one who generates the multipartite state) performs a measurement on one or more modes of the multipartite state, and the outcome is used as the reference to generate the conference key.
Reverse Reconciliation: One of the remote users' measurements is used as the reference to generate the conference key.
Entanglement-in-the-Middle: All modes are sent to the remote users without the dealer retaining any part of the state. Any remote user's measurement can serve as the reference, analogous to protocols where entanglement is distributed without a central party.
After generating the conference key, the conditional measurement data can still be used to establish bipartite keys post-QCKA between the remaining users. Additionally, we consider generating bipartite keys directly between the dealer and the users without involving QCKA.

Our results demonstrate that the proposed protocol using the dual-rail cluster state and quotient graph state not only outperforms the GHZ/W states generated with the same level of squeezing for QCKA using direct reconciliation, achieving positive key generation under a wider range of channel conditions, but also significantly outperforms them for generating bipartite keys post-QCKA across all methods (direct reconciliation, reverse reconciliation, and entanglement-in-the-middle). Moreover, when compared to a downstream access network utilizing two-mode squeezed vacuum states with equivalent squeezing, our protocol achieves superior performance in generating bipartite keys without involving QCKA.

Recognizing the importance of practical implementations, we extend our analysis to consider real world conditions. We examine the performance of our protocol in the finite-size regime, which is crucial when the number of signals exchanged is limited. Our finite-size analysis provides insights into the key rates achievable under realistic conditions, accounting for statistical fluctuations due to finite sample sizes.

Additionally, we consider the impact of using impure squeezed states for generating the multipartite entangled states. In practical scenarios, experimental imperfections lead to squeezed states that are not perfectly pure \cite{Hsieh2022}, affecting the overall performance of quantum protocols. By incorporating impure squeezed states into our analysis, we provide a more realistic assessment of our protocol's robustness against imperfections in state preparation.

This study highlights the importance of quotient graph states in CV quantum systems for state engineering in quantum networks, offering improved key rates and robustness against losses, noise, and experimental imperfections. We also introduce a more accurate method for estimating bipartite key rates in downstream access networks, enhancing the evaluation of CV system's key generation capabilities.

This paper is structured as follows: In Section~\ref{classical part}, we delve into the complexities of classical information processing for multiple users. Section~\ref{sec:security} presents the security analysis of a general multipartite state distributed among trusted users and provides their key rate expressions. In Section~\ref{sec:tripartite state}
, we discuss the generation of various tripartite states from squeezed states, including the impact of impurity. In Section~\ref{sec:dual}, we introduce techniques for utilizing dual-rail cluster states in various cryptographic protocols. In Section~\ref{sec:qkd_protocols} we introduce and analyze various multi-user cryptographic protocols. In Section~\ref{sec:results} we summarize the major results of our analysis. Finally, in Section~\ref{sec:conclusion}
, we review the outcomes of our research and discuss potential avenues for future work.

\section{Multivariate Gaussian Correlations and Multi-User QKD}\label{classical part}

This section simplifies the discussion of multivariate Gaussian correlations by focusing on three users, \(A\), \(B\), and \(C\), although the principles discussed can be extended to larger networks. To analyze the inter-user relationships and their implications for QKD, it is useful to classify the types of correlations that may occur. In the context of error correction and key generation, the structure of these correlations profoundly influences the strategies we must employ.

In what follows, we delineate three primary correlation scenarios that capture the essence of these interactions:
\subsection*{Correlation Scenarios}
\begin{description}
    \item[Symmetric Correlations :] All users share equal correlations. Any user can disclose error correction information (syndrome) to enable the others to derive the same raw key, facilitating a shared conference key after privacy amplification.

    \item[Centralized Correlations :] User A is more strongly correlated with users B and C than they are with each other, allowing for:
    \begin{itemize}
        \item \textbf{Conference Key Agreement (CKA):} User A shares the syndrome, enabling users B and C to establish a conference key.
        \item \textbf{Independent Key Generation:} Users B and C transmit their syndromes to A, allowing A to establish separate keys with each of them.
    \end{itemize}

    \item[Induced Correlations :] Initially uncorrelated users become correlated if a user or dealer, already correlated with them, shares the necessary syndrome for error correction.
\end{description}

Elaborating on the second and third scenarios, consider three users, specifically $A,B$ and $C$. $A$ aims to generate a key with $B$ and $C$, which can be accomplished through two separate techniques. The first technique requires $A$ to convey the essential information for error correction to $B$ and $C$. Conversely, $B$ and $C$ could forward the necessary information for error correction to $A$. In the initial scenario, the keys generated will be perfectly correlated as both $B$ and $C$ use the data possessed by $A$ to generate a key. However, in the second scenario, the situation gets more complicated if $B$ and $C$ are correlated, resulting in a potential partial correlation of the keys. In the event of correlations, $B$ and $C$ cannot be treated separately, as the mutual information $I(A:B,C) \neq I(A:B)+I(A:C)$, but rather $I(A:B,C)=I(A:B)+I(A:C|B)=I(A:C)+I(A:B|C)$. The mutual information is defined as $I(i:j)=H(i)-H(i|j)=H(j)-H(j|i)$, and the conditional mutual information $I(i:j|k)=H(i|k)-H(i|j,k)$. $H(.)$ represents the Shannon entropy.
When dealing with multiple correlated users, it becomes essential to ascertain the optimal syndrome length, that needs to be transferred for successful error correction. In the given example, $B$ and $C$ must jointly transmit an amount of information quantified as $H(BC|A)$ to $A$. The restriction lies in the collective syndrome delivered by $B$ and $C$. The individual syndromes of $B$, say $S_B$, and $C$, say $S_C$, can exceed in length, provided that $H(S_B,S_C)=H(BC|A)$.

When $B$ and $C$ exhibit no correlation, i.e., $I(B:C)=0$, this does not necessarily mean that the conditional mutual information $I(B:C|A)=0$. If both $B$ and $C$ are correlated with $A$, then disclosing $A$, or ideally, having $A$ communicate the syndromes to $B$ and $C$ for error correction and to establish a conference key among $A$, $B$, and $C$, will induce a correlation in the conditional data between $B$ and $C$. This, in turn, enables them to generate a key among themselves. It should be noted that the sequence in which keys are generated influences the nature of the keys that can be produced. In the discussed case, a 1D graph facilitated the creation of a conference key among $A$, $B$, and $C$. Subsequently, $B$ and $C$ are able to establish a key between themselves. Alternatively, if $B$ or $C$ instead of $A$ transmits the syndrome to $A$, they are capable of creating separate keys with $A$.  In a more general setting, the conditional covariance between $B$ and $C$, represented as $\langle B_A C_A\rangle = \mathcal{C}_3 - \frac{\mathcal{C}_1 \mathcal{C}_2}{V}$, where $B_A$ and $C_A$ denote the conditional data held by $B$ and $C$,  disclosing $A$ diminishes the correlation between $B$ and $C$ if $Sgn(\mathcal{C}_1 \mathcal{C}_2)/V = Sgn(\mathcal{C}_3)$, eliminates it when $\mathcal{C}_1 \mathcal{C}_2/V = \mathcal{C}_3$, and enhances it under other circumstances. Here, $\mathcal{C}_1=\langle A B\rangle$, $\mathcal{C}_2=\langle A C\rangle$, $\mathcal{C}_3=\langle B C\rangle$, and $V=\langle A^2\rangle=\langle B^2\rangle=\langle C^2\rangle$.

\begin{figure}[htbp]
    \centering
    \includegraphics[width=0.8\linewidth]{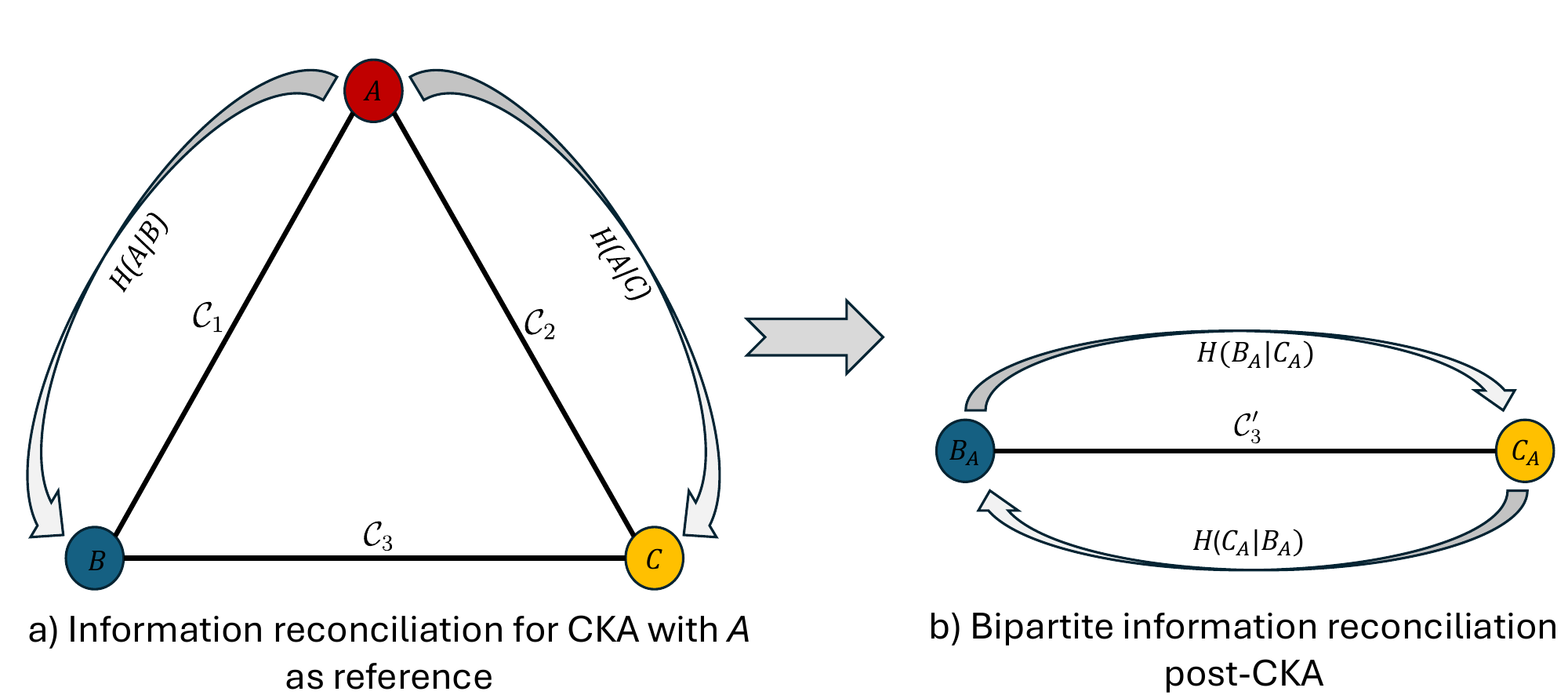}
    \caption{(a) Information reconciliation for Conference Key Agreement (CKA) with user \(A\) as the reference; (b) Bipartite information reconciliation post-CKA.}
    \label{fig:conference_scheme}
\end{figure}

\begin{figure}[htbp]
    \centering
    \includegraphics[width=0.6\linewidth]{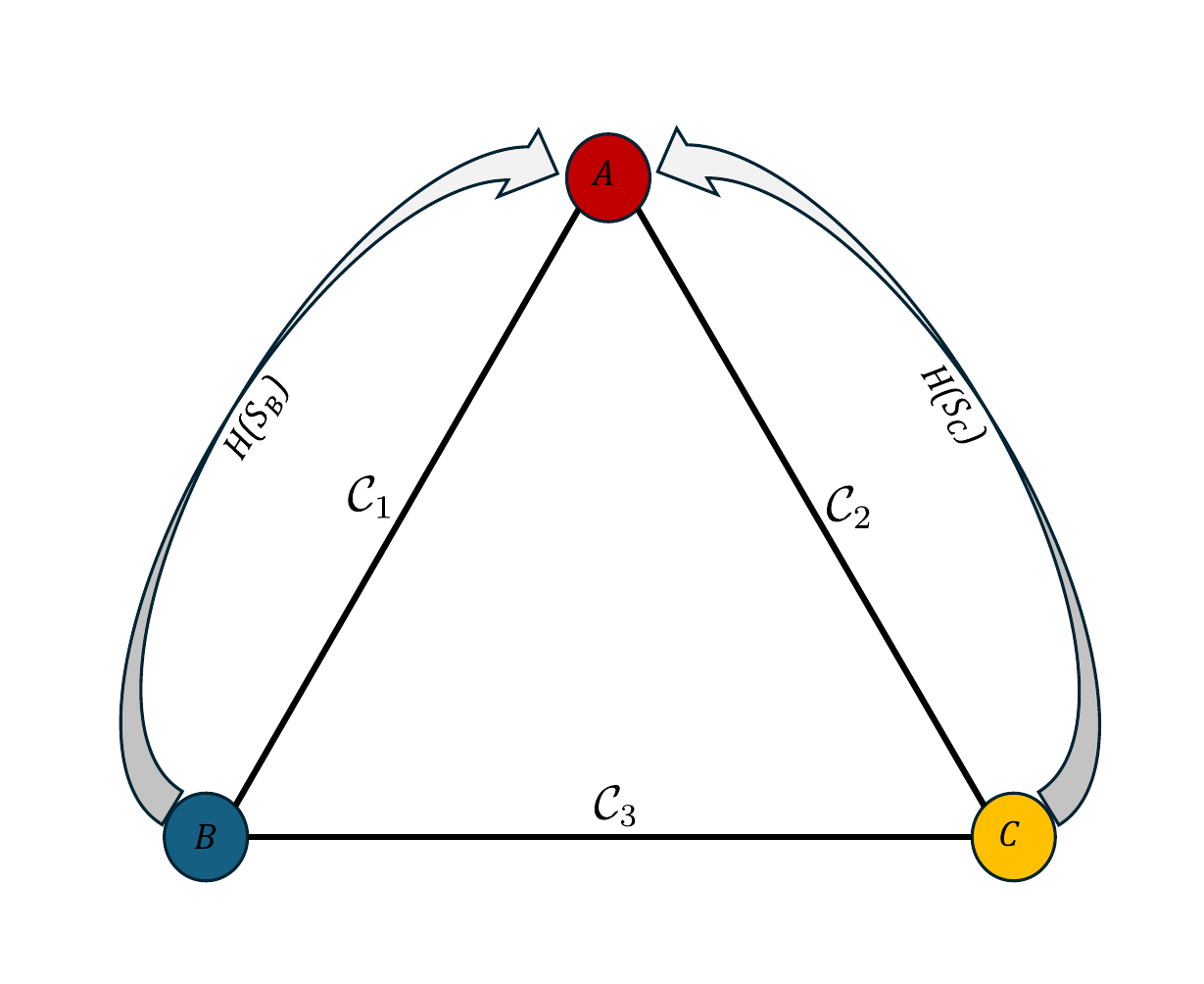}
    \caption{Information reconciliation for independent bipartite key generation between user $A$ and users $B$ and $C$. User $A$ establishes separate keys with $B$ and with $C$, each sending a syndrome $S_B$ or $S_C$ to $A$. The quantities $H(S_B)$ and $H(S_C)$ denote the entropies of the communicated syndromes, and their optimal joint entropy satisfies $H(S_B,S_C)=H(BC\mid A)$.}

    \label{fig:independent_scheme}
\end{figure}

Fig.~\ref{fig:conference_scheme} illustrates the information reconciliation process for CKA with \(A\) as the reference. In this scenario, after reconciliation, \(B\) and \(C\) can also extract a bipartite key (see parts (a) and (b) of Fig.~\ref{fig:conference_scheme}). Fig.~\ref{fig:independent_scheme} depicts the information reconciliation process for independent bipartite key generation, where user \(A\) establishes separate keys with \(B\) and \(C\).

\section{Security analysis}\label{sec:security}
\subsection{Holevo bound}\label{sec:holevo}

In the context of CV-QKD, the Holevo bound represents the upper limit on the information accessible to an eavesdropper about the measurement outcomes of the parties whose data are used to generate a secure key. To determine this bound, we assume that the eavesdropper, $E$, purifies the state of the legitimate parties, denoted as $A$, $B$, and $C$ (we consider three parties for simplicity, but the underlying principles can be extended to more parties and modes) \cite{GaussianQI,GarciaPatron2006}. This entanglement leads us to posit that the entire system $ABCE$ is in a pure state. Consequently, the von Neumann entropy of the combined system, $S(ABCE)$, is zero.

Before proceeding, we must clarify our notation. Throughout this work, the term $S(I|J)$ denotes the entropy of subsystem $I$ evaluated in the state, $\rho_{I|J}$, which results \textit{after} a measurement on subsystem $J$. The entropy is calculated as $S(I|J) := S(\rho_{I|J}) = -\text{Tr}[\rho_{I|J} \log_2 \rho_{I|J}]$.

This definition should not be confused with the standard conditional von Neumann entropy, which is an algebraic property of the initial (pre-measurement) state $\rho_{IJ}$. The standard definition, which we denote as $S(I|J)_{\text{std}}$, is given by $S(I|J)_{\text{std}} = S(IJ) - S(J)$, or explicitly:
\begin{equation}
    S(I|J)_{\text{std}} = -\text{Tr}[\rho_{IJ} \log_2 \rho_{IJ}] + \text{Tr}[\rho_J \log_2 \rho_J]
\end{equation}

Crucially, our notation $S(I|J)$ refers to the entropy of a state produced by a physical measurement process, while the standard definition does not.

With this notational framework, we can establish the relationships needed to bound Eve's information. Since the overall state of $ABCE$ is pure, the entropy of any subsystem is equal to that of its complement; thus, $S(ABC) = S(E)$. Furthermore, a measurement on subsystem $A$ collapses the remaining system $BCE$ into a new pure state. For any tripartite pure state, the entropy of a subsystem is equal to that of its complement. This physical principle directly leads to the equality $S(BC|A) = S(E|A)$, where both quantities are entropies of the respective subsystems in the post-measurement state conditioned on the outcome at $A$.

Eve's knowledge of the measurement outcome of $A$ is upper-bounded by the Holevo quantity $\chi_A$, which is the initial entropy of Eve's system reduced by the entropy remaining after Alice's measurement. This can be written as $\chi_A = S(E) - S(E|A)$. Using the relations established above, this bound becomes:
\begin{equation}
   \chi_{A} = S(ABC) - S(BC|A).
\end{equation}

This framework can be extended to sequential key generation. Here, the analysis of a subsequent key between $B$ and $C$ is performed on the conditional state $\rho_{BC|A}$. This approach, where the protocol conditions on $A$'s data and the security proof assumes this data is known to all parties including Eve, is a deliberate choice made for two critical security reasons. First, the assumption that Eve has access to $A$'s data allows us to establish the most conservative (worst-case) upper bound on her information for the subsequent key. Second, by generating the key only from the data corresponding to the conditional state $\rho_{BC|A}$, we ensure the new key is statistically independent (uncorrelated) of $A$'s original data. This prevents vulnerabilities that could arise from cross-key correlations.

It is important to note that this conditioning does not require a sequential quantum measurement process. The protocol imposes no specific temporal order on the measurements; each party simply measures their respective quantum state upon receipt, and the 'conditioning' is performed entirely during classical post-processing. As will be detailed in the key rates analysis (Sec.~\ref{sec:keyrates}), this classical step requires at least one of the parties ($B$ or $C$) to have knowledge of $A$'s data. Therefore, the analysis of this stage is not an examination of a hypothetical scenario, but a direct assessment of the cryptographic resource available to the remaining parties under this robust security model. For a key generated between $B$ and another party using this state, the information leaked to Eve is bounded by quantities that depend on the entropy of the full remaining system, $S(BC|A)$, reduced by the entropy of the system left after $B$'s measurement. The bounds are thus given by:
\begin{equation}
    \chi_{B|A} = S(BC|A) - S(C|AB)
\end{equation}

and similarly, for a key generated by $C$:
\begin{equation}
    \chi_{C|A} = S(BC|A) - S(B|AC).
\end{equation}

We consider Gaussian states that are fully described by their first moments (mean quadrature values) and second moments, captured in the covariance matrix \( \Sigma \). This matrix is defined as:

\begin{equation}
    \Sigma_{ij} = \frac{1}{2} \left\langle \hat{x}_i \hat{x}_j + \hat{x}_j \hat{x}_i \right\rangle - \left\langle \hat{x}_i \right\rangle \left\langle \hat{x}_j \right\rangle,
\end{equation}

where \( \left\langle \cdot \right\rangle \) denotes expectation values. Diagonal elements of the matrix indicate the variance of each quadrature, representing their uncertainties, while off-diagonal elements represent correlations between different quadratures.

The von Neumann entropy of these states is derived from the symplectic eigenvalues \( \nu_i \) of the covariance matrix \cite{GaussianQI} and is given by:

\begin{equation}
    S= \sum_{i} \left[ \left(\nu_i + \frac{1}{2}\right) \log \left(\nu_i + \frac{1}{2}\right) - \left(\nu_i - \frac{1}{2}\right) \log \left(\nu_i - \frac{1}{2}\right) \right].
\end{equation}

Similarly, conditional entropies \( S(p|q) \) can be derived from the corresponding conditional covariance matrices. The conditional covariance matrix \(\Sigma_{p|q}\) is expressed as:

\begin{equation}
    \Sigma_{p|q} = \Sigma_p - \Sigma_{pq} (X \Sigma_q X)^{MP} \Sigma_{pq}^T,
    \label{eq:condmat}
\end{equation}

where \(\Sigma_p\) denotes the covariance matrix of the unmeasured modes of the multipartite state, \(\Sigma_q\) represents the covariance matrix of the measured modes, \(\Sigma_{pq}\) is the covariance matrix between the measured and unmeasured modes, and \(X\) is defined as \(Diag(1,0)\) for measurements in the x-quadrature and $MP$ in the Moore-Penrose inverse.

The overall covariance matrix \(\Sigma\) of the system, incorporating both measured and unmeasured modes, is structured as follows:

\begin{equation}
    \Sigma = \begin{pmatrix}
        \Sigma_p & \Sigma_{pq} \\
        \Sigma_{pq}^T & \Sigma_q
    \end{pmatrix}
\end{equation}

Throughout this work, we assume homodyne detection of the \(x\)-quadrature unless stated otherwise. For all graph states studied here, except the GHZ/W state, the quadratures are symmetric in phase space, meaning the variances of the \(x\)- and \(p\)-quadratures are equal. As a result, the choice of measurement quadrature does not affect the key rate or security analysis. In the case of the GHZ/W state, our construction yields lower variance in the \(x\)-quadrature, making it the favorable choice for measurement. We also considered heterodyne detection as an alternative, but it was consistently found to be suboptimal compared to homodyne detection across all scenarios analyzed. To maintain clarity and avoid redundancy, all references to measurement in this paper should be understood as \(x\)-quadrature homodyne detection.

\subsection{Key Rates}\label{sec:keyrates}

Having addressed the complexities of multi-user QKD in the contexts of information reconciliation and the estimation of the Holevo bound, we are now positioned to establish expressions for various key rates \cite{DevetakWinter2004,devetak2005distillation,GaussianQI}. For a three-user system, three distinct types of keys can be generated:

\begin{enumerate}
    \item Conference Key Agreement (CKA)
    \item Bipartite Key Generation post-CKA
    \item Independent Bipartite Key Generation
\end{enumerate}

\subsubsection*{Conference Key Agreement (CKA)}

Conference Key Agreement can be realized among the users if there exists at least one user who is correlated with the remaining users. The amount of information (syndrome) that needs to be communicated to the rest of the users is determined by the lowest mutual information between this central user and the other users. The conference key rate can be expressed as:
\begin{equation}\label{eq:conf}
K^{i}(i:j:k) = \beta \min[I(i:j), I(i:k)] - \chi_i
\end{equation}
$\beta$ is the reconciliation efficiency \cite{beta} and the superscript signifies the reference data of the user $i$ used to generate the key. The efficiency $\beta$ is a crucial practical parameter that accounts for the fact that real-world error correction protocols, used by the parties to reconcile their correlated data into an identical string, are not perfectly efficient and have a communication overhead.

\subsubsection*{Bipartite Keys Post-CKA}

Once a CKA with user $i$ as the reference is complete, the conditional data possessed by the remaining two users, $j$ and $k$, can be utilized to generate a new, independent key. Following a successful CKA, both users $j$ and $k$ now possess the measurement outcomes of user $i$. This allows either party to act as the reference for the subsequent key generation. A secure key can be generated from this conditional resource provided that the conditional mutual information between $j$ and $k$, $I(j:k|i)$, is greater than the information Eve can obtain about the raw key.

To determine the Holevo bound for this subsequent stage, we apply the same security principles as before. We pessimistically assume that Eve also knows the measurement outcomes of $i$. The security analysis is then performed on the conditional state $\rho_{jk|i}$, where we posit that Eve holds a purification of this state. Following the arguments presented in Sec.~\ref{sec:holevo}, we can then calculate the bound on Eve's information for this specific conditional resource. The key rate for this scenario depends on which user acts as the reference for error correction. If user $j$ is the reference, the key rate is given by:
\begin{equation}
K^{j}(j:k|i) = \beta I(j:k|i) - \chi_{j|i}
\end{equation}
If, instead, $k$ is used as the reference, the key rate is:
\begin{equation}
K^{k}(j:k|i) = \beta I(j:k|i) - \chi_{k|i}
\end{equation}

\subsubsection*{Independent Bipartite Keys}

A user can generate independent bipartite keys with the users with whom it is correlated, provided the mutual information of this user (say $i$) with the rest of the users is higher than the mutual information between the other users (ideally zero). The key rates are:
\begin{equation}
K^{j}(i:j) = \beta I(i:j) - \chi_j
\end{equation}
 and $K^{k}(i:k|j)$. In the scenario where the mutual information between the other users is zero, the equation simplifies to $K^{k}(i:k)$. Notice that even when \( j \) and \( k \) are identical, the order of key generation influences the key rates. This sequential dependence is critical. After user $i$ establishes a key with user $j$ (corresponding to the rate $K^j(i:j)$), user $i$ now possesses the measurement outcomes of user $j$ after successful error correction. This knowledge enables the subsequent step. For the key rate $K^k(i:k|j)$, where user $k$ provides the reference data, key generation is possible because user $i$ has the necessary conditional information (its own data and $j$'s data) to perform error correction and privacy amplification. However, alternate reconciliation with key rate $K^i(i:k|j)$ would not be described by this framework, as it would require user $k$ to perform error correction and privacy amplification, for which it would need access to $j$'s measurement outcomes—information they do not have. It is more useful to consider the sum of independent keys that \( i \) can generate with other users. The actual key rates with each user depend on the information disclosed for error correction, specifically the syndrome length. The sum of independent keys is given by:
\begin{equation}
\begin{split}
    K^{j,k}(i\!:\!jk) &= H(j,k) - \zeta H(j,k|i) - \chi_{jk} \\
    &= K^j(i\!:\!j) + K^k(i\!:\!k|j)
\end{split}
\end{equation}

As discussed in Sec.~\ref{classical part}, \( H(j,k|i) \) represents the optimal amount of information that \( j \) and \( k \) need to communicate to \( i \) to facilitate error correction and $\zeta\ge 1$ accounts for inefficient information reconciliation. For the symmetric case, the independent key rates can be computed as follows:
\begin{align}
    K(i\!:\!j) = K(i\!:\!k)
    &= \tfrac{1}{2} \Big[ H(j,k) - \zeta H(\mathbf{S}_j, \mathbf{S}_k) \nonumber \\
    &\quad - S(ijk) + S(i \mid jk) \Big]
\end{align}

Here, \( \bold{S}_j \) and \( \bold{S}_k \) are the syndromes with entropies \( H(\bold{S}_j) = H(\bold{S}_k) \). In this article, we focus on the conditions under which a positive key rate can be achieved, and therefore, we emphasize the sum of independent keys rather than individual key rates.

\subsection{Finite Size Analysis}

In practical implementations, the number of exchanged quantum signals is inherently finite. This finite data sample introduces statistical fluctuations that can significantly impact the security and performance of the QKD protocol. Unlike the asymptotic regime, discussed in the previous section, where it is assumed that an infinite number of signals are exchanged, allowing statistical uncertainties to vanish. In the finite-size regime, the key rate is affected by the statistical estimation of channel parameters.

The order in which error correction and parameter estimation are performed affects the finite-size key rate expression \cite{Leverrier2015}. When error correction is performed \textit{after} parameter estimation, some transmitted signals are disclosed to estimate the channel parameters. Let $N$ be the total number of signals exchanged, and $n$ be the number of signals used for key generation after discarding $N - n$ signals for parameter estimation. The finite-size key rate $K$ is given by:

\begin{equation}\label{eq:finite}
K = \frac{n}{N} \left[ K_{\infty}(t^{\text{low}}, V_{\epsilon}^{\text{up}}) - \Delta(n) \right]
\end{equation}

Alternatively, when error correction is performed \textit{before} parameter estimation, all measurements can be used for both parameter estimation and key generation, allowing for better estimates of the channel parameters since no data is discarded \cite{Leverrier2015,Walenta2014}. In this case, the finite-size key rate $K$ is:

\begin{equation}\label{eq:opt_finite_key}
K = K_{\infty}(t^{\text{low}}, V_{\epsilon}^{\text{up}}) - \Delta(n)
\end{equation}

In both expressions, \( K_{\infty}(t^{\text{low}}, V_{\epsilon}^{\text{up}}) \) is the asymptotic key rate calculated using conservative estimates of the channel transmission \( t \) and excess noise \( V_{\epsilon}= t\epsilon \) to account for statistical fluctuations due to the finite sample size. These adjusted parameters are:

\begin{align}
t^{\text{low}} &= t - 6.5\, \sigma_{t} \\
V_{\epsilon}^{\text{up}} &= V_{\epsilon} + 6.5\, \sigma_{\epsilon}
\end{align}

where \( \sigma_{t} \) and \( \sigma_{\epsilon} \) are the standard deviations of the estimators of \( t\) and \( V_{\epsilon} \), respectively (see Appendix~\ref{sec:PE}). The factor \( 6.5 \) corresponds to a confidence level ensuring an error probability of \( 10^{-10} \). The finite-size correction term \( \Delta(n) \) \cite{finite1}accounts for the statistical fluctuations in the mutual information:

\begin{equation}
\Delta(n) = 7 \sqrt{\dfrac{\log_2(2 \times 10^{10})}{n}}
\end{equation}

with \( n \) being the number of signals used for key generation.

\section{Generation of Tripartite States}\label{sec:tripartite state}

The generation of multipartite states in CV systems can be efficiently realized by manipulating squeezed light states through beam splitters \cite{threemodecv,cvthree}. This method allows us to explore different configurations of squeezed states and beam splitters that result in distinct tripartite states. We define a general structure of the covariance matrix as follows:
\begin{equation}\label{eq:cm}
\begin{pmatrix}
    V_A & C_{AB} & C_{AC} \\
    C_{AB} & V_B & C_{BC} \\
    C_{AC} & C_{BC} & V_C
\end{pmatrix}
\end{equation}
Here, $V_i$ represents the variance of mode $i = A, B, C$, and $C_{ij}$ denotes the covariance between the quadratures of modes $i$ and $j$.

We consider two configurations: the symmetric GHZ-like state and the downstream access network (DAN). The GHZ-like state is of interest because it has maximum correlation equally distributed among all modes. The DAN configuration is appealing because it can be easily implemented by substituting the two-mode squeezed vacuum with either modulated squeezed or coherent states, and it offers an advantage in producing bipartite keys.

\begin{figure}
    \centering
    \includegraphics[width=\columnwidth]{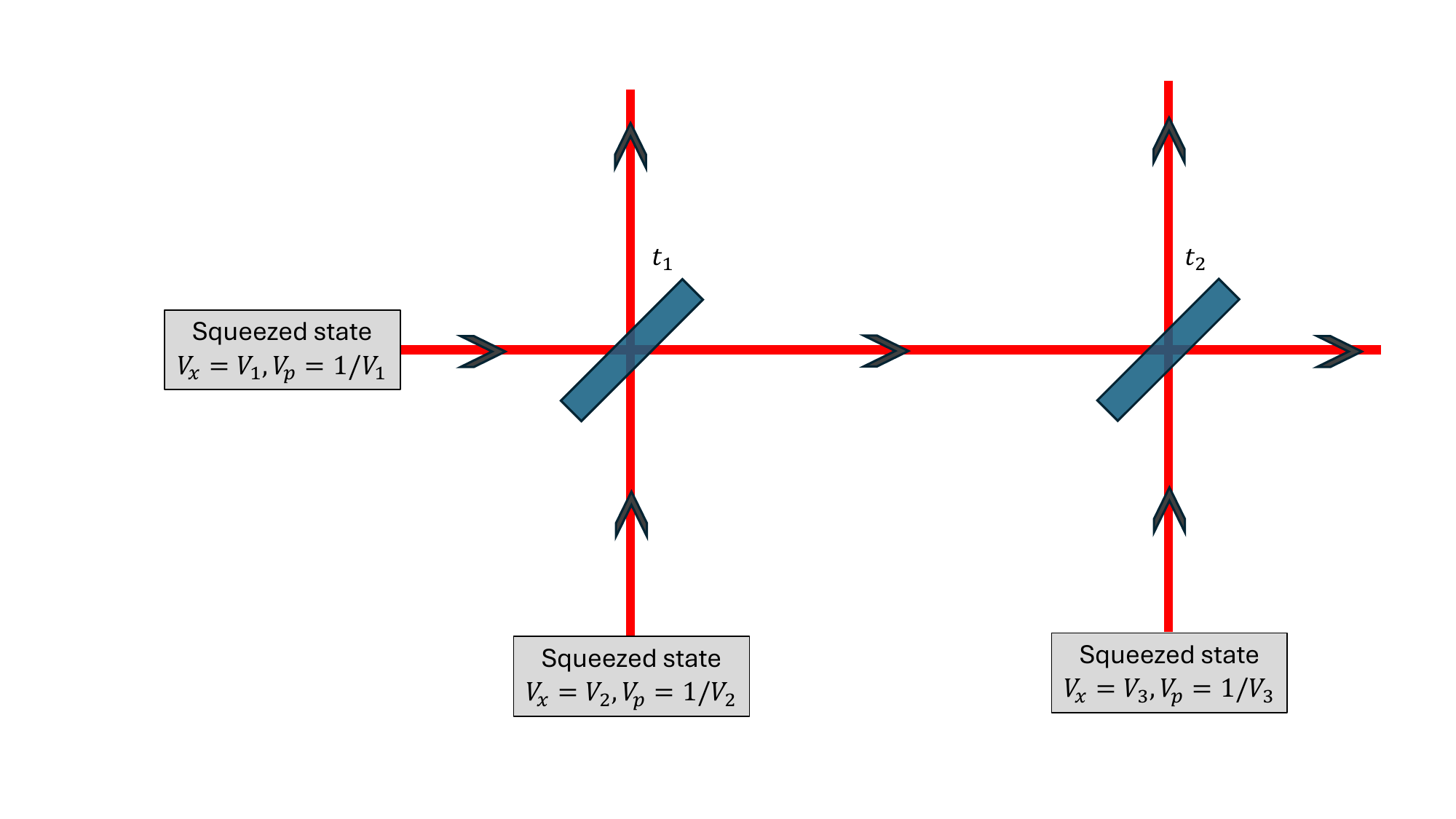}
    \caption{Schematic diagram for generating tripartite states}
    \label{fig:tripartite}
\end{figure}

In the first configuration, referred to as the \textbf{Symmetric GHZ-like state}, three squeezed states with $x$-quadrature variances \(V_1 = V\), \(V_2 = \frac{1}{V}\), and \(V_3 = \frac{1}{V}\) are directed into two beam splitters with transmissions \(t_1 = \frac{2}{3}\) and \(t_2 = \frac{1}{2}\) (see Fig.~\ref{fig:tripartite}). This setup yields a symmetric covariance matrix where
\begin{equation}
    V_A = V_B = V_C = \text{Diag}\left[\frac{1+2V^2}{3V},\ \frac{2+V^2}{3V}\right],
\end{equation}
and the covariances between modes are given by
\begin{equation}
    C_{AB} = -C_{AC} = C_{BC} = \text{Diag}\left[\frac{V^2 - 1}{3V},\ \frac{1 - V^2}{3V}\right].
\end{equation}

This state exhibits maximum correlation equally among all modes, analogous to the GHZ state in discrete-variable systems.

The second configuration is based on the \textbf{Downstream Access Network (DAN)} approach, which is a cost-effective setup for generating tripartite states. In this scheme, two squeezed states with $x$-quadrature variances \(V_1 = V\) and \(V_2 = \frac{1}{V}\), along with a vacuum mode \(V_3 = 1\), are combined using two beam splitters, with transmissions \(t_1 = t_2 = \frac{1}{2}\). This results in a covariance matrix where
\begin{equation}
\begin{aligned}
V_A &= \mathrm{Diag}\!\left[\tfrac{1+V^2}{2V},\, \tfrac{1+V^2}{2V}\right], \\[6pt]
V_B &= V_C = \mathrm{Diag}\!\left[\tfrac{(1+V)^2}{4V},\, \tfrac{(1+V)^2}{4V}\right], \\[6pt]
C_{AB} &= -C_{AC} = \mathrm{Diag}\!\left[\tfrac{V^2 - 1}{2\sqrt{2}V},\, \tfrac{1 - V^2}{2\sqrt{2}V}\right], \\[6pt]
C_{BC} &= \mathrm{Diag}\!\left[-\tfrac{(V - 1)^2}{4V},\, -\tfrac{(1 - V)^2}{4V}\right].
\end{aligned}
\end{equation}

This configuration simplifies implementation and reduces costs compared to other schemes, as the two-mode squeezed vacuum can be replaced with a modulated squeezed state.

\subsubsection{Effect of Impurity in Squeezed States}

In practical implementations, squeezed states are often not pure due to technical imperfections and decoherence, which introduce additional noise in the anti-squeezed quadrature, which usually is proportional to the degree of squeezing \cite{Hsieh2022}. We characterize this impurity by an additional noise parameter \(V_N\), such that if the variance of the squeezed quadrature is \(V\), the variance of the anti-squeezed quadrature becomes \(V_N + 1/V\).

To account for the impurity, we modify the covariance matrices of the input squeezed states. For the GHZ-like state configuration, the covariance matrices of the input squeezed states become
\begin{equation}
\begin{aligned}
V_1 &= \mathrm{Diag}\!\left[V,\, V_N + \tfrac{1}{V}\right], \\[6pt]
V_2 &= V_3 = \mathrm{Diag}\!\left[V_N + \tfrac{1}{V},\, V \right].
\end{aligned}
\end{equation}

Similarly, for the DAN configuration, the covariance matrices are
\begin{equation}
\begin{aligned}
V_1 &= \mathrm{Diag}\!\left[V,\, V_N + \tfrac{1}{V}\right], \\[6pt]
V_2 &= \mathrm{Diag}\!\left[V_N + \tfrac{1}{V},\, V\right], \\[6pt]
V_3 &= \mathrm{Diag}\!\left[1,\, 1\right].
\end{aligned}
\end{equation}

Using these modified input variances, we can compute the resulting covariance matrices after the beam splitter transformations, taking into account the impurity.

For the \textbf{GHZ-like state} with impure squeezed states, the variances and covariances become
\begin{equation}
\begin{aligned}
V_A &= V_B = V_C
= \mathrm{Diag}\!\left[\tfrac{1+2V^2+V V_N}{3V},\, \tfrac{2+V^2+2V V_N}{3V}\right],
\end{aligned}
\end{equation}

and the covariances between modes are given by

\begin{equation}
\begin{aligned}
C_{AB} &= -C_{AC} = C_{BC}
= \mathrm{Diag}\!\left[\tfrac{V^2 - 1 - V V_N}{3V},\, \tfrac{1 - V^2 + V V_N}{3V}\right].
\end{aligned}
\end{equation}

For the \textbf{DAN} configuration, the variances and covariances are
\begin{equation}
\begin{aligned}
V_A &= \mathrm{Diag}\!\left[\tfrac{1+V^2+V V_N}{2V},\, \tfrac{1+V^2+V V_N}{2V}\right], \\[6pt]
V_B &= V_C = \mathrm{Diag}\!\left[\tfrac{(1+V)^2+V V_N}{4V},\, \tfrac{(1+V)^2+V V_N}{4V}\right], \\[6pt]
C_{AB} &= -C_{AC} = \mathrm{Diag}\!\left[\tfrac{V^2 - 1 - V V_N}{2\sqrt{2}V},\, \tfrac{1 - V^2 + V V_N}{2\sqrt{2}V}\right], \\[6pt]
C_{BC} &= \mathrm{Diag}\!\left[-\tfrac{(V - 1)^2+V V_N}{4V},\, -\tfrac{(1 - V)^2+V V_N}{4V}\right].
\end{aligned}
\end{equation}

By incorporating the impurity parameter \( V_N \), we can examine the effect of non-ideal squeezing on the generated tripartite states. The presence of additional noise increases the variances of the anti-squeezed quadratures, thereby impacting the correlations among the modes. This noise must be carefully accounted for to ensure the viability of cryptographic tasks. We assume that this preparation noise is trusted and does not provide information to eavesdroppers.

To address this, we introduce an additional mode for each squeezed state to account for the noise, ensuring that the two modes together form a pure bipartite state. Specifically, we consider the impure squeezed state as one of the outputs of a beam splitter fed by two pure squeezed states with x-quadrature variances \(\Gamma_1\) and \(\Gamma_2\) at the input. The impure squeezed state is then related to the variances of the input pure squeezed states as follows:
\begin{equation}
\begin{aligned}
\Gamma_1 &= V - \frac{V^3 V_N}{\sqrt{V^3 V_N \bigl(1 + V^3 V_N\bigr)}}, \\[6pt]
\Gamma_2 &= V + \frac{V^3 V_N}{\sqrt{V^3 V_N \bigl(1 + V^3 V_N\bigr)}}.
\end{aligned}
\end{equation}

\section{Dual-rail cluster state for QKD}\label{sec:dual}
In this analysis, we propose a distribution strategy for the modes of a dual-rail cluster state in a multi-node QKD network. Unlike a finite graph defined by a discrete set of nodes, the dual-rail cluster state represents an open-ended, repeating lattice of interconnected modes. Each mode is typically linked to four others, creating an extensive and complex structure that is difficult to handle as it is.

To make this infinitely repeating configuration more tractable, we introduce a coloring scheme inspired by graph theory terminology. By assigning colors to modes and then grouping all modes of the same color together, we form what we call a quotient graph state. This quotient graph serves as a finite, homomorphic image of the original dual-rail lattice, preserving key connectivity patterns while reducing its complexity. Although we draw on terms like “coloring” and “quotient graph” from graph theory, our approach focuses on using these concepts to describe and manage the grouping process, rather than relying on advanced graph-theoretic techniques.

The resulting quotient graph state provides a simplified framework that retains the essential structural features of the dual-rail cluster state. In doing so, it creates a more manageable platform for implementing robust and scalable continuous-variable QKD protocols, ensuring that the underlying entanglement resources can be effectively harnessed in practical quantum communication networks.

\begin{figure}
     \centering

     \begin{subfigure}[b]{\columnwidth}
         \centering
         \includegraphics[width=\columnwidth]{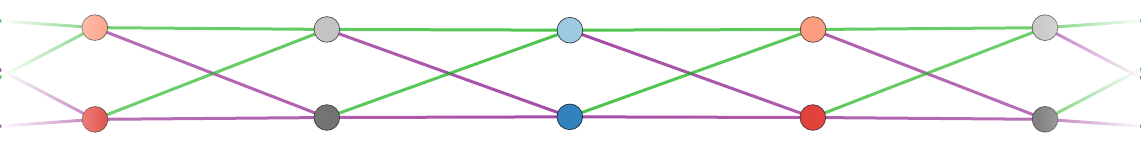}
         \caption{Dual rail cluster state}
         \label{fig:dual}
     \end{subfigure}
     \vfill
     \begin{subfigure}[b]{0.45\columnwidth}
         \centering
         \includegraphics[width=\columnwidth]{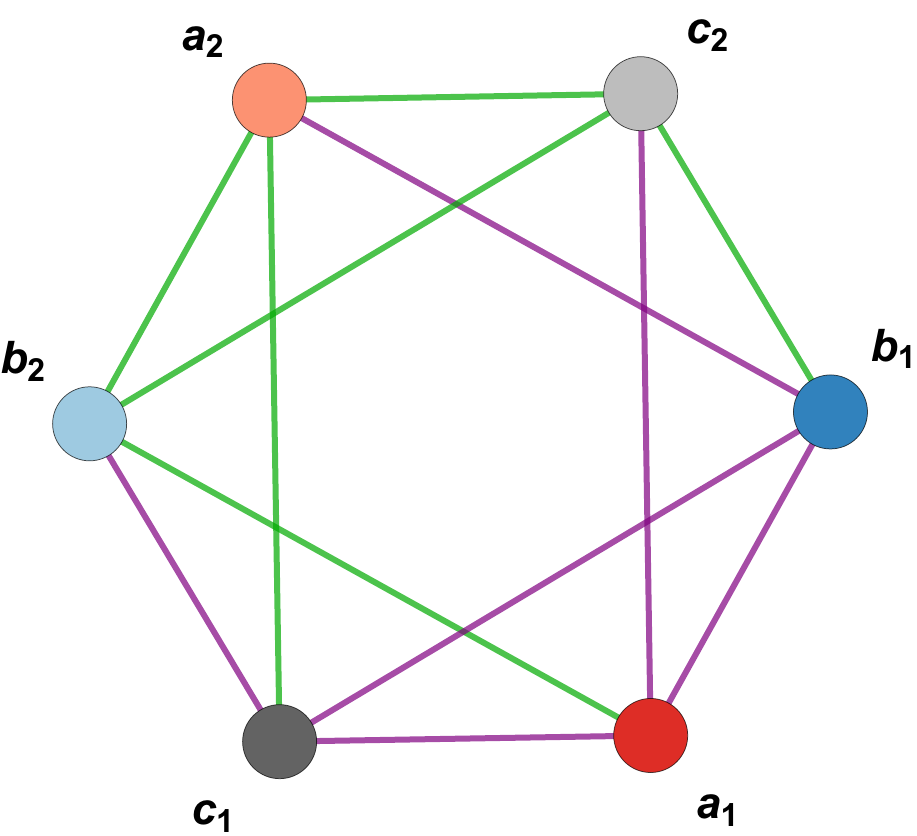}
         \caption{6 mode graph state}
         \label{fig:6graph}
     \end{subfigure}
        \caption{(a) Shows the coloring scheme of dual rail cluster state to obtain the six mode graph state (b). The edges in the graph illustrates the correlations between modes, with edges in two colors indicating the sign of the correlation. Edges of one color represent positive correlations, while edges of the other color represent negative correlations. }
        \label{fig:eqivalent graphs}
\end{figure}

The optimal homomorphic graph corresponds to a six-mode graph state, achievable through a coloring strategy detailed in Fig.~\ref{fig:eqivalent graphs} . The covariance matrix of this six-mode graph state is (refer to Sec.~\ref{sec:dual to six} for derivation):
\begin{equation}\label{eq:6modecm}
\begin{pmatrix}
   \boldsymbol{V} & \boldsymbol{0} & \boldsymbol{C}&-\boldsymbol{C}&\boldsymbol{C}&\boldsymbol{C} \\
   \boldsymbol{0} & \boldsymbol{V} &\boldsymbol{C} &-\boldsymbol{C}&-\boldsymbol{C}&-\boldsymbol{C} \\
   \boldsymbol{C}&\boldsymbol{C}&\boldsymbol{V} &\boldsymbol{0} &\boldsymbol{C}&-\boldsymbol{C}\\
   -\boldsymbol{C}&-\boldsymbol{C}&\boldsymbol{0}&\boldsymbol{V}&\boldsymbol{C}&-\boldsymbol{C}\\
   \boldsymbol{C}&-\boldsymbol{C}&\boldsymbol{C}&\boldsymbol{C}&\boldsymbol{V}&\boldsymbol{0}\\
   \boldsymbol{C}&-\boldsymbol{C}&-\boldsymbol{C}&-\boldsymbol{C}&\boldsymbol{0}&\boldsymbol{V}
\end{pmatrix}
\end{equation}
where the diagonal covariance matrix \(\boldsymbol{V}\) and the off-diagonal covariance matrix \(\boldsymbol{C}\) are defined as follows for pure squeezed states:
\begin{equation}\label{eq: covariance1}
   \boldsymbol{V}= \begin{pmatrix}
    \frac{1+V^2}{2V} & 0 \\
    0 & \frac{1+V^2}{2V}
\end{pmatrix}, \quad \boldsymbol{C}=\begin{pmatrix}
    \frac{1-V^2}{4V} & 0 \\
    0 & \frac{V^2-1}{4V}
\end{pmatrix}.
\end{equation}

$V$ is the variance of the squeezed quadrature of the squeezed state. For impure squeezed states, as described in the previous section, the impurity parameter \(V_N\) modifies these matrices to:
\begin{equation}\label{eq:covariance2}
\begin{aligned}
\boldsymbol{V} &=
\begin{pmatrix}
    \tfrac{1+V^2+V_N}{2V} & 0 \\[6pt]
    0 & \tfrac{1+V^2+V_N}{2V}
\end{pmatrix}, \\[10pt]
\boldsymbol{C} &=
\begin{pmatrix}
    \tfrac{1 - V^2 + V V_N}{4V} & 0 \\[6pt]
    0 & \tfrac{V^2 - 1 - V V_N}{4V}
\end{pmatrix}.
\end{aligned}
\end{equation}

This six-mode graph state provides a versatile platform for various QKD strategies. We present optimal methods for generating a three-user conference key and independent bipartite keys, which we compare to the three-mode tripartite states discussed in the previous section.

\subsection{Distribution strategies for six-mode graph state}

In our investigation of six-mode graph states for QKD, we explored a wide array of strategies for distributing the quantum states among users. However, due to space constraints and to focus on the most impactful designs, here we present only the strategies that proved to be the most noteworthy in terms of operational performance and security. We categorize the strategies based on whether the dealer also participates as a user or acts solely as a distributor.

\subsubsection*{Dealer as a Participant}

\begin{enumerate}
  \item \textbf{Distribution 1 (Fig.~\ref{fig:dealer4})}: The dealer sends mode \(b_1\) and mode \(c_1\) to two distant users through a channel characterized by channel transmission $t$ and channel excess noise $\epsilon$ transforming the modes to $\Tilde{b}_1$ and $\Tilde{c}_1$, with covariance matrix $\Tilde{\boldsymbol{V}}=t (\boldsymbol{V}+\epsilon \mathbb{I}-\mathbb{I})+\mathbb{I}$, $\mathbb{I}$ here is $2\times2$ identity matrix. The dealer retains the rest of the modes, \(a_1\),\(a_2\), \(b_2\), and \(c_2\).

  \item \textbf{Distribution 2 (Fig.~\ref{fig:dealer2})}: The dealer sends modes $b_1$ and $b_2$ to one user and modes $c_1$ and $c_2$ to the other user. The dealer retains modes \(a_1\) and \(a_2\).
\end{enumerate}

\subsubsection*{Dealer as a Distributor}

\begin{figure}[h]
    \centering
    \begin{centering}
        \begin{subfigure}{0.45\columnwidth}
            \centering
            \includegraphics[width=\columnwidth]{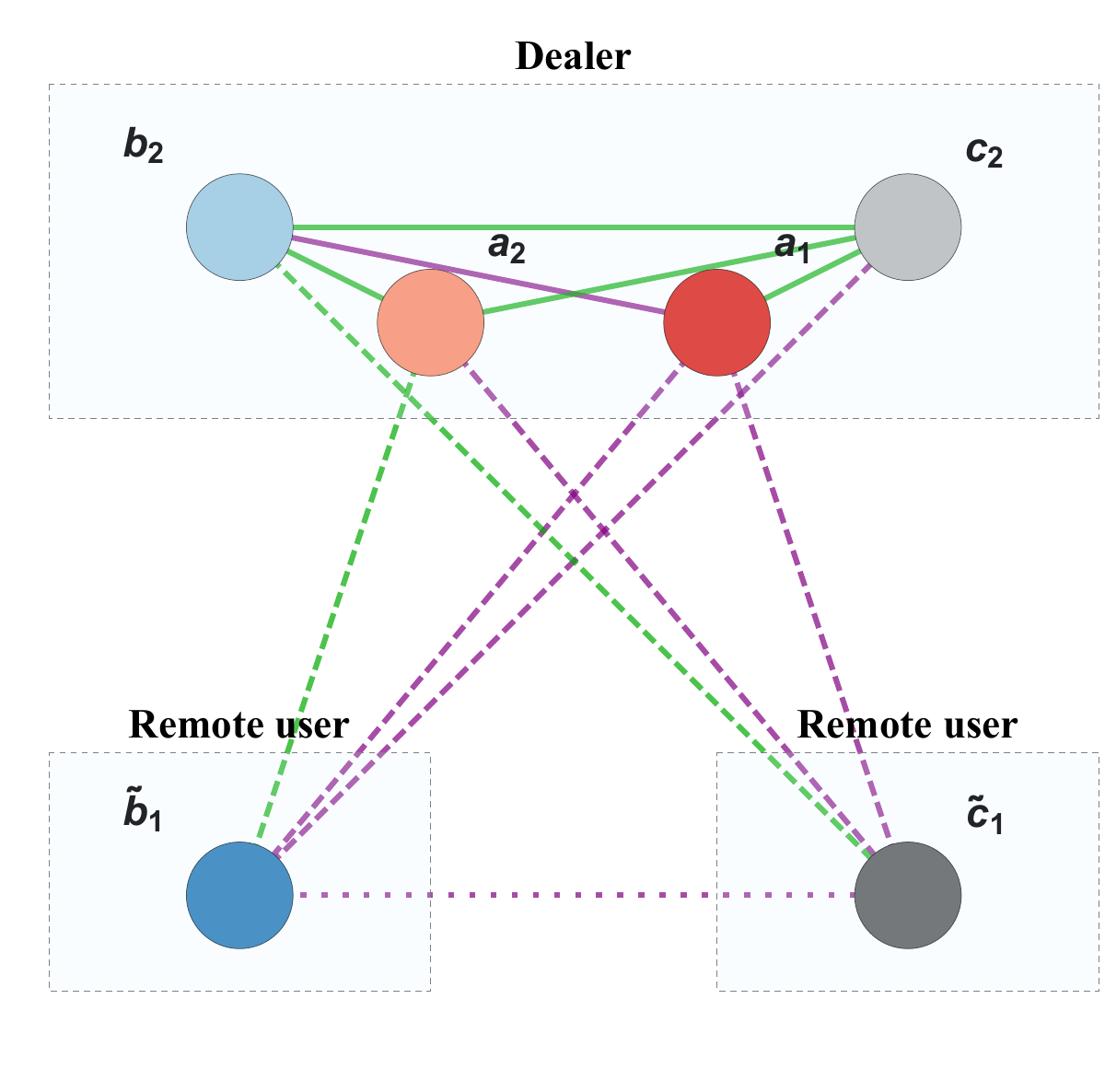}
            \caption{Distribution 1}\label{fig:dealer4}
        \end{subfigure}
        \hspace{0.05\columnwidth} 
        \begin{subfigure}{0.45\columnwidth}
            \centering
            \includegraphics[width=\columnwidth]{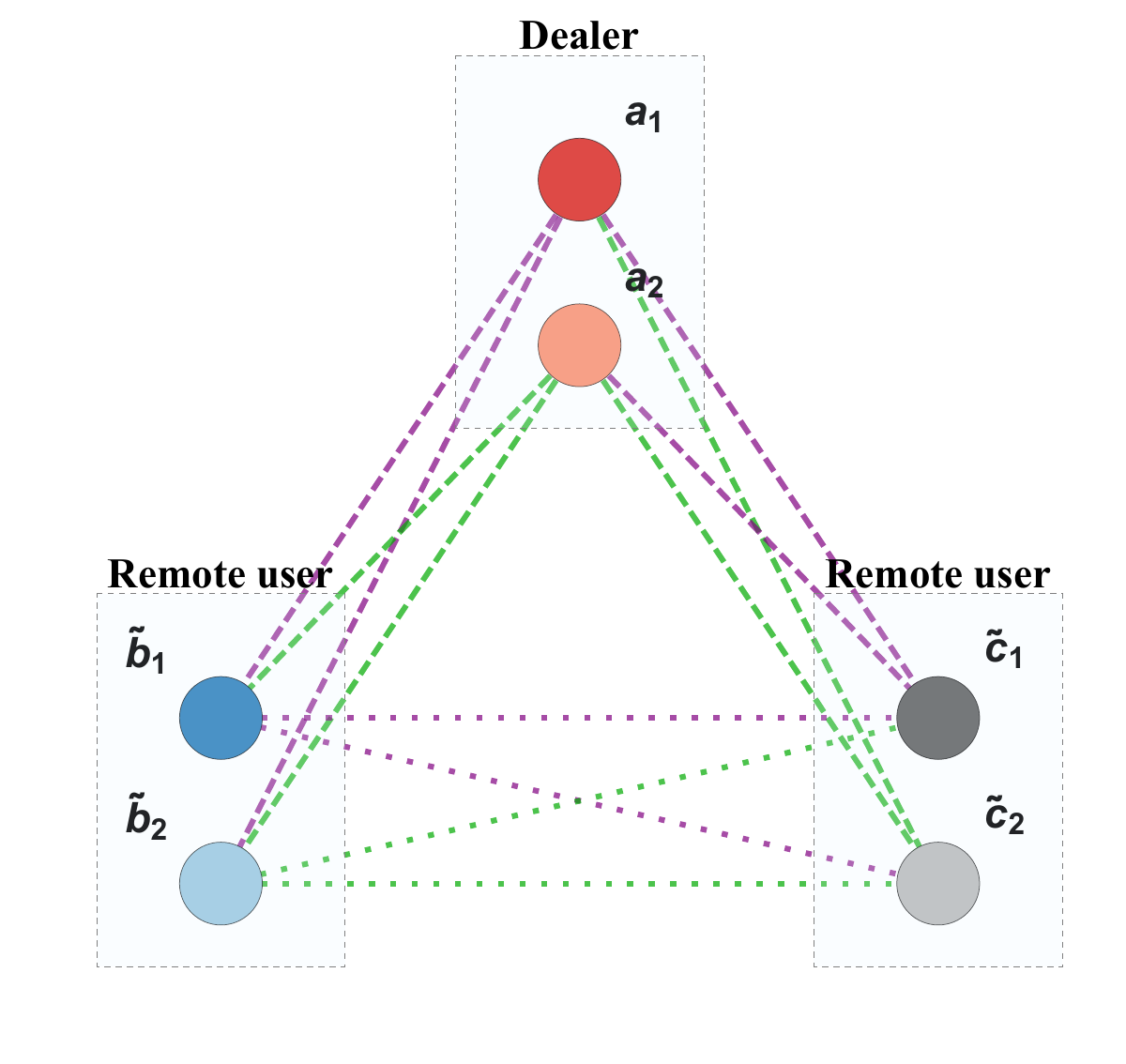}
            \caption{Distribution 2}\label{fig:dealer2}
        \end{subfigure}
    \end{centering}

    \vspace{0.5cm} 

    \begin{centering}
        \begin{subfigure}{0.45\columnwidth}
            \centering
            \includegraphics[width=\columnwidth]{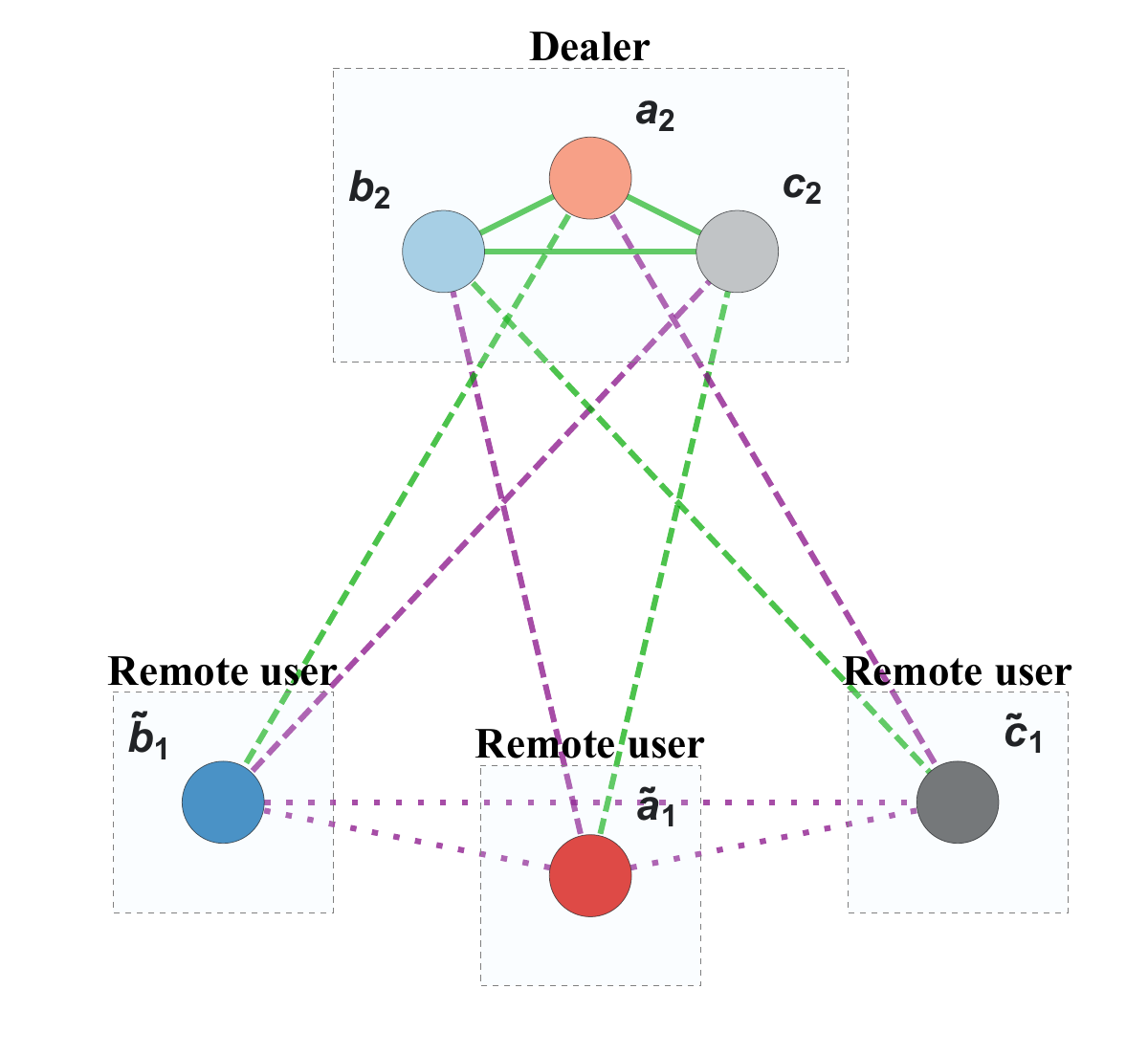}
            \caption{Distribution 3}\label{fig:dealer3}
        \end{subfigure}
        \hspace{0.05\columnwidth} 
        \begin{subfigure}{0.45\columnwidth}
            \centering
            \includegraphics[width=\columnwidth]{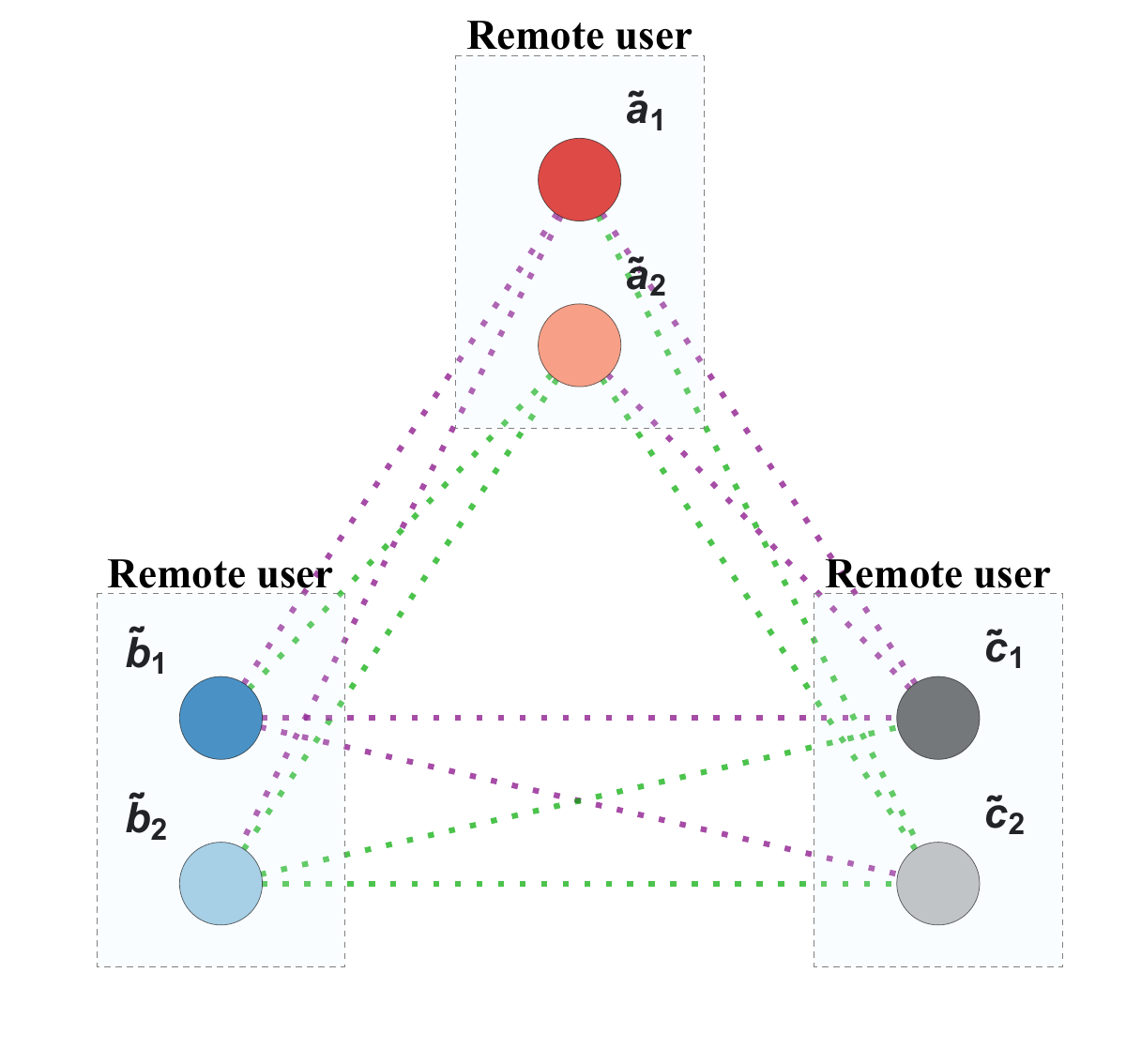}
            \caption{Distribution 4}\label{fig:dealer0}
        \end{subfigure}
    \end{centering}

    \caption{The graphs depict various methods for distributing a six-mode graph state among three users. Dashed lines indicate that the correlations (i.e., covariance between the modes $\boldsymbol{C}$) have been reduced by a factor of $\sqrt{t}$, while dotted lines indicate a reduction by a factor of $t$.}
\end{figure}

\begin{enumerate}
  \item \textbf{Distribution 3 (Fig.~\ref{fig:dealer3})}: Modes \(a_1\), \(b_1\), and \(c_1\) are distributed to three separate users, the rest are retained by the dealer.

  \item \textbf{Distribution 4 (Fig.~\ref{fig:dealer0})}: All the modes are sent through the channel to three remote users, one user gets \(a_1\) and \(a_2\), another receives \(b_1\) and \(b_2\), and the third \(c_1\) and \(c_2\).
\end{enumerate}

Each of these strategies offers distinct advantages, and the choice of strategy should consider the operational context and security requirements of the QKD system.

\section{Quantum conference key distribution protocols}
\label{sec:qkd_protocols}

Quantum conference key distribution protocols like standard two-user QKD protocols can be broadly categorized into three categories, based on the role of the dealer and the method of key reconciliation. Understanding these protocols is essential for evaluating the security and efficiency of multi-user quantum communication systems. In our analysis, we assume that the channel parameters are symmetric, meaning that all users have identical channel conditions.

\subsection*{Direct Reconciliation (DR)}
In the direct reconciliation protocol, the dealer retains one or more modes of the multipartite state they prepare and uses their measurement outcomes as references for key generation. The remote parties receive the necessary information and perform the required error correction.

For the GHZ state described in Sec.~\ref{sec:tripartite state}, the covariance matrix given in Eq.~\ref{eq:cm} transforms to:

\begin{equation}\label{eq:cm2t}
\begin{pmatrix}
V_A & \sqrt{t}\, C_{AB} & \sqrt{t}\, C_{AC} \\
\sqrt{t}\, C_{AB} & t(V_B + \epsilon - 1) + 1 & t\, C_{BC} \\
\sqrt{t}\, C_{AC} & t\, C_{BC} & t(V_C + \epsilon - 1) + 1
\end{pmatrix}
\end{equation}

The key rate for this approach is given by \( k^{A}(A:\tilde{B}:\tilde{C}) \) (refer to Eq.~\ref{eq:conf}), where the notation \( \tilde{i} \) indicates that mode \( i \) has been transmitted through a quantum channel.

We explore two distribution strategies for a six-mode graph state intended for three-user conference key generation via direct reconciliation, as illustrated in Figs.~\ref{fig:dealer4} and \ref{fig:dealer2}.

There are multiple methods for conference key generation based on \emph{Distribution 1} (Fig.~\ref{fig:dealer4}). Among these, the key \( k^{a_1}(a_1:\tilde{b}_1:\tilde{c}_1\,|\,a_2) \) proves most robust against channel losses and noise in the asymptotic regime when reconciliation efficiency is perfect (\(\beta=1\)); see Fig.~\ref{fig:dr_asy}. Note that \( a_1 \) and \( a_2 \) are not interchangeable due to the correlation type (sign) with the remote user's modes. Depending on the channel parameters, leveraging measurements from multiple modes as references can further enhance the conference key rate.

From Fig.~\ref{fig:dr_asy}, we observe that under finite squeezing, the GHZ state is not the optimal multipartite resource for maximizing distance in conference key generation with direct reconciliation under ideal conditions. Among the configurations tested, \emph{Distribution 1} achieves the greatest distances, particularly under high excess channel noise (\(0.08 < \epsilon < 0.5\) SNU). Although trusted noise does not hinder the performance of \emph{Distribution 1} or GHZ-based protocols in the asymptotic regime (and may even assist them), it offers no similar advantage to \emph{Distribution 2}, yet it does not significantly degrade its performance either.

In the finite-size regime, for \(n=10^7\) measurements and imperfect reconciliation (\(\beta=0.95\)), \emph{Distribution 2} (Fig.~\ref{fig:dealer2}) with key rate \( k^{a_1}(a_1:\tilde{b}_1\tilde{b}_2:\tilde{c}_1\tilde{c}_2\,|\,a_2) \) enables key generation over greater distances for channel noise $0<\epsilon<0.27 SNU$; see Fig.~\ref{fig:dr_finite}. This is in contrast to what we observe in the asymptotic regime where \emph{Distribution 1} enables key generation for the largest distances. This is due to the fact that \emph{Distribution 2} enables the users and the dealer to use four correlations as opposed to three for \emph{Distribution 1} to enable better estimation of the channel parameters (see Sec.~\ref{sec:PE}).

\subsection*{Reverse Reconciliation (RR)}
Contrary to direct reconciliation, reverse reconciliation involves a remote user serving as the reference. This user sends error correction information to the dealer and other participants, facilitating key alignment based on their reference measurements.

The conference key rate for tripartite states under reverse reconciliation is given by \( k^{\tilde{B}}(A:\tilde{B}:\tilde{C}) \).

For the six-mode graph state with \emph{Distribution 1} (Fig.~\ref{fig:dealer4}), we find that the optimal key rate for achieving higher tolerance to channel parameters under reverse reconciliation is
\[
k^{\tilde{b}_1}(a_1:\tilde{b}_1:\tilde{c}_1\,|\,b_2 a_2 c_2).
\]

For Distribution 2 (Fig.~\ref{fig:dealer2}), the optimal conference key is
\[
k^{\tilde{b}_1}(a_1 a_2:\tilde{b}_1 \tilde{b}_2:\tilde{c}_1 \tilde{c}_2\,|\,\tilde{b}_2).
\]

From Fig.~\ref{fig:rr_asy}, it is apparent that the GHZ state generates positive keys under a broader range of channel conditions. Comparable performance is achieved with \emph{Distribution 1}. The influence of trusted noise for reverse reconciliation is the same as for direct reconciliation, it shows little to no effects on protocols with \emph{Distribution 2} but benefits protocols with \emph{Distribution 1} and the GHZ state.

In the finite size regime; see Fig.~\ref{fig:rr_finite}, the trend remains the same, with the GHZ state enabling the largest distances followed by \emph{Distribution 1} and \emph{Distribution 2}. Though the difference is not significant.

\subsection*{Entanglement in the middle (Mid)}

In this protocol, the dealer prepares entangled states and distributes them among the users but does not have access to the conference key being generated.

The conference key rate for tripartite states in the entanglement-in-the-middle protocol is given by \( k^{\tilde{A}}(\tilde{A}:\tilde{B}:\tilde{C}) \).

For the six-mode graph state with \emph{Distribution 3} (Fig.~\ref{fig:dealer3}), the dealer discloses the measurement outcomes of all of their modes, effectively disconnecting themselves from the conditional data held by the other users. This leads to the conference key rate between the users being
\[
k^{\tilde{a}_1}(\tilde{a}_1:\tilde{b}_1:\tilde{c}_1\,|\,a_2 b_2 c_2).
\]

It should be noted that instead of disclosing the measurement outcomes, the dealer can generate three sets of conference keys with any two remote users, with conference key rates \( k^{a_2}(a_2:\tilde{b}_1:\tilde{c}_1) \), \( k^{b_2}(b_2:\tilde{a}_1:\tilde{c}_1\,|\,a_2) \), and \( k^{c_2}(c_2:\tilde{a}_1:\tilde{b}_1\,|\,a_2 b_2) \). However, the tolerance to channel noise and loss for the conference key \( k^{b_2}(b_2:\tilde{a}_1:\tilde{c}_1\,|\,a_2) \) was found to be very low.

For the \emph{Distribution 4} (Fig.~\ref{fig:dealer2}), the optimal conference key is
\[
k^{\tilde{a}_1 \tilde{a}_2}(\tilde{a}_1 \tilde{a}_2:\tilde{b}_1 \tilde{b}_2:\tilde{c}_1 \tilde{c}_2).
\]

\begin{figure}[t]
    \centering

    \subfloat[]{%
        \includegraphics[width=0.4\columnwidth]{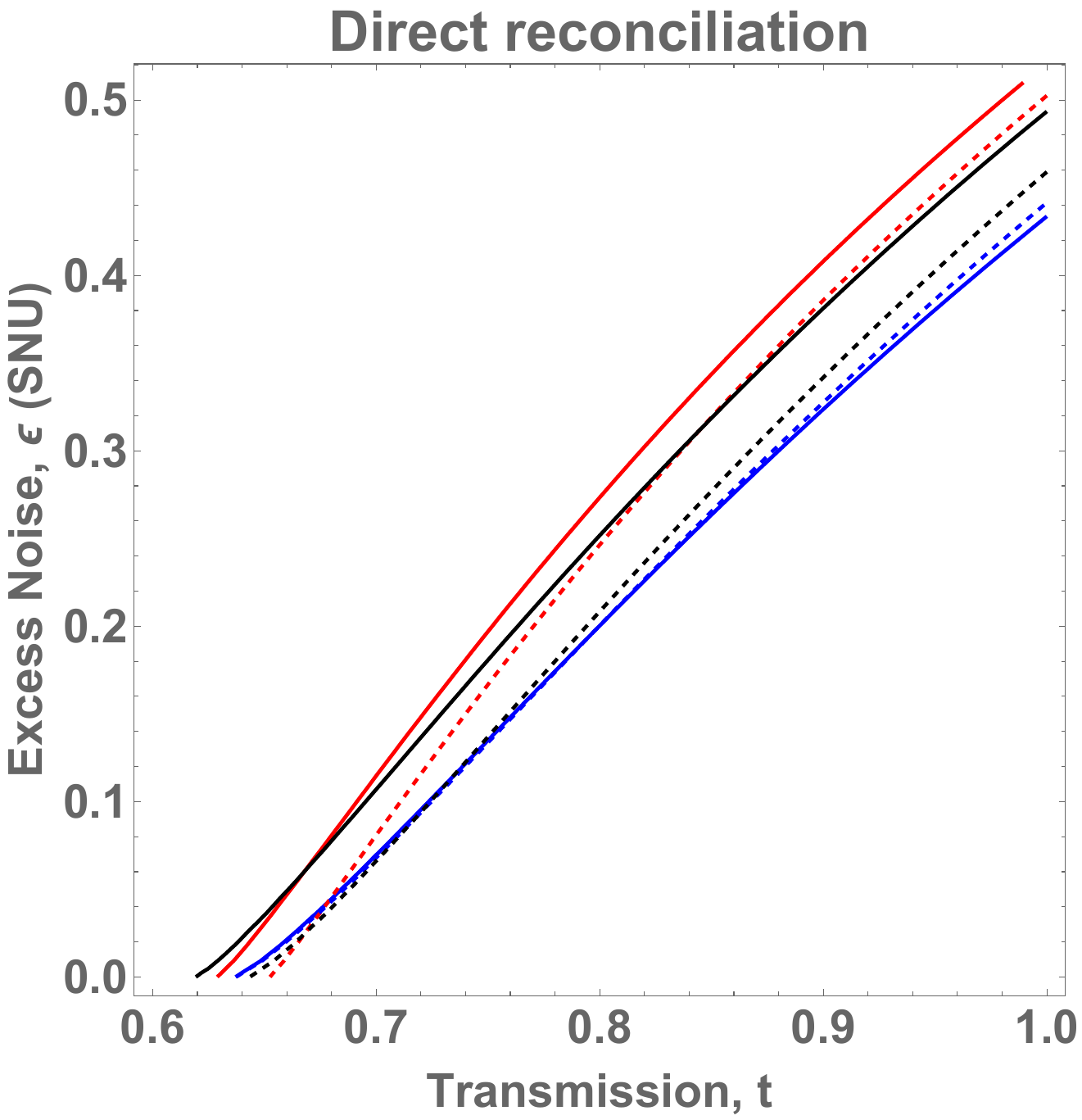}%
        \label{fig:dr_asy}%
    }\hspace{0.02\columnwidth}
    \subfloat[]{%
        \includegraphics[width=0.4\columnwidth]{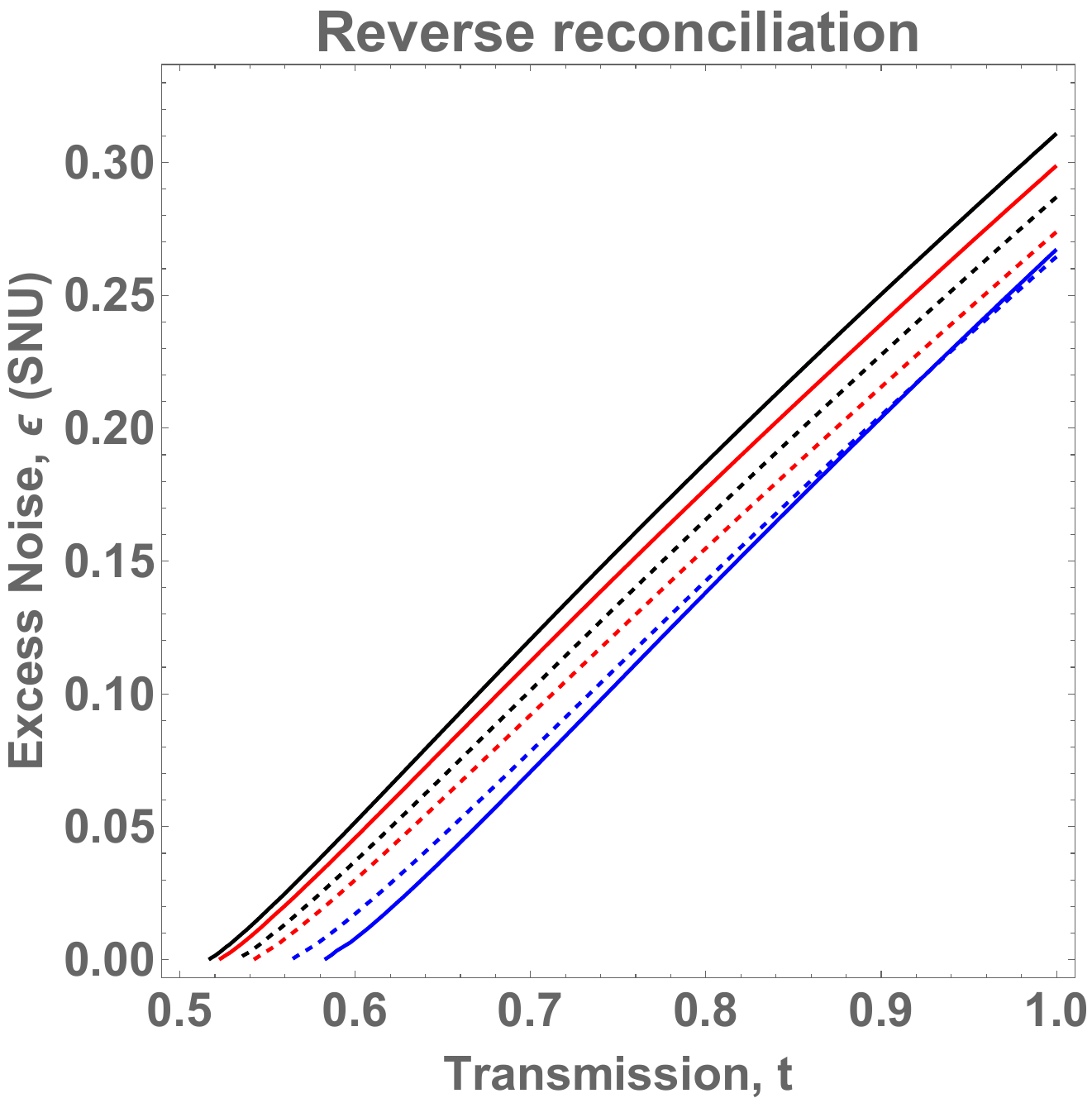}%
        \label{fig:rr_asy}%
    }\hspace{0.02\columnwidth}
    \subfloat[]{%
        \includegraphics[width=0.4\columnwidth]{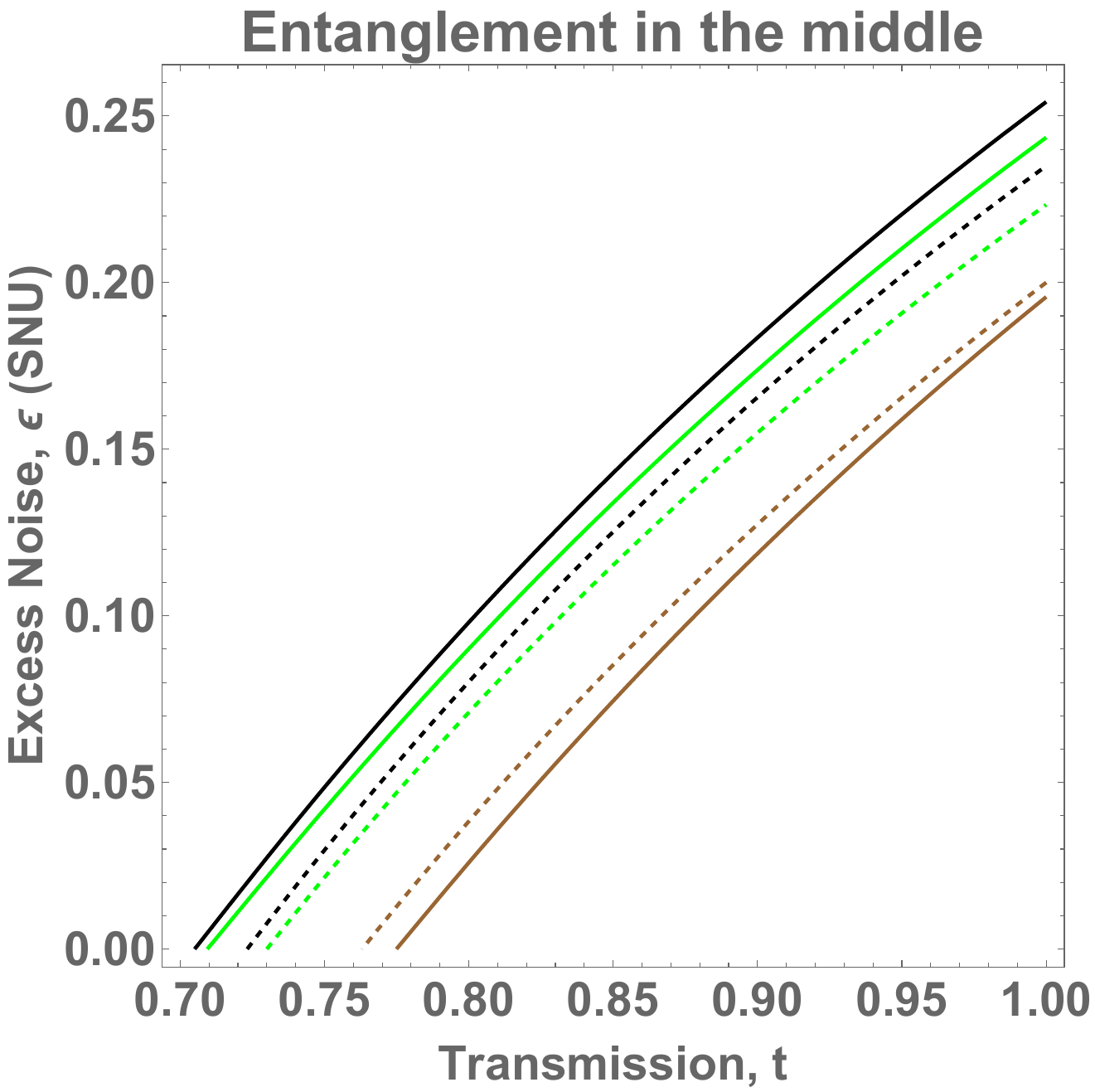}%
        \label{fig:mid_asy}%
    }

    \caption{These plots illustrate the maximum tolerable channel losses and noise in asymptotic regime, key generation is possible only for channel parameters lying below the plotted curves. The blue curve corresponds to \emph{Distribution 2}, the red to \emph{Distribution 1}, the green to \emph{Distribution 3}, the brown to \emph{Distribution 4}, and the black to the GHZ state. Dashed lines represent pure squeezed states, while solid lines indicate impure but trusted squeezed states with $V_N = 10$. The reconciliation efficiency is $\beta = 1$, and the variance of the squeezed quadrature is $V = 0.1$ SNU. }
    \label{fig:ckaasymtotic}
\end{figure}

\begin{figure}[t]
    \centering

    \subfloat[]{%
        \includegraphics[width=0.4\columnwidth]{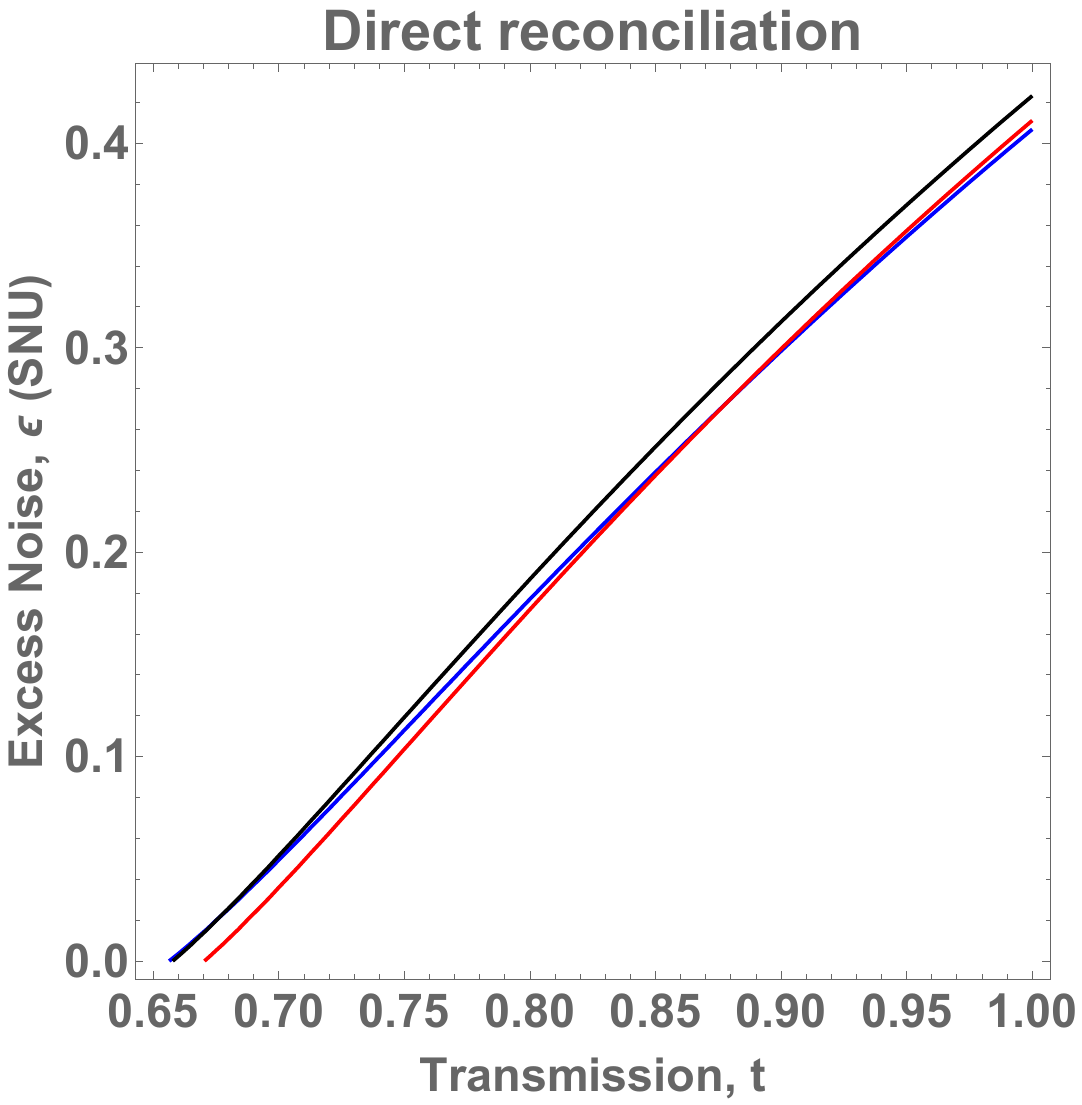}%
        \label{fig:dr_finite}%
    }\hspace{0.02\columnwidth}
    \subfloat[]{%
        \includegraphics[width=0.4\columnwidth]{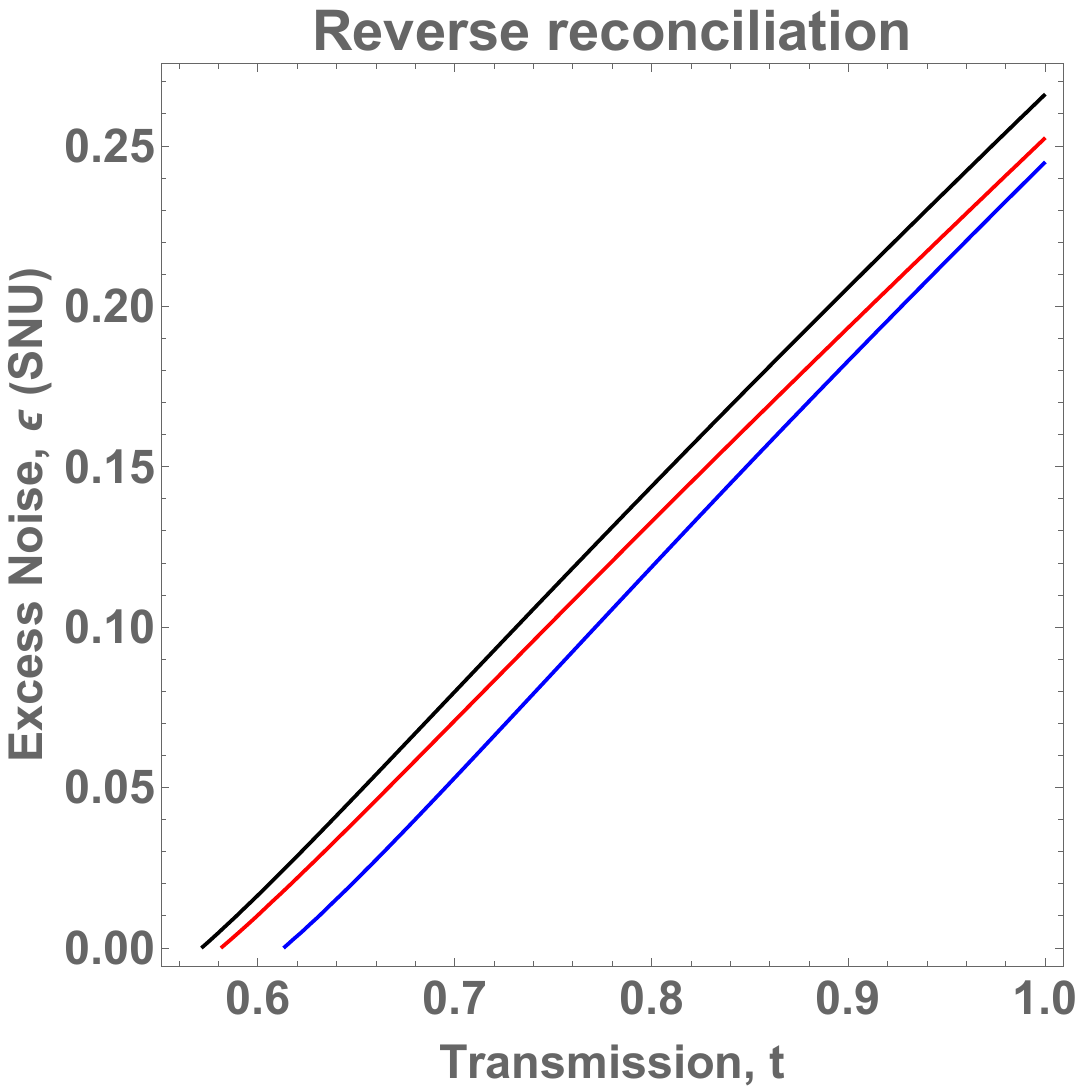}%
        \label{fig:rr_finite}%
    }\hspace{0.02\columnwidth}
    \subfloat[]{%
        \includegraphics[width=0.4\columnwidth]{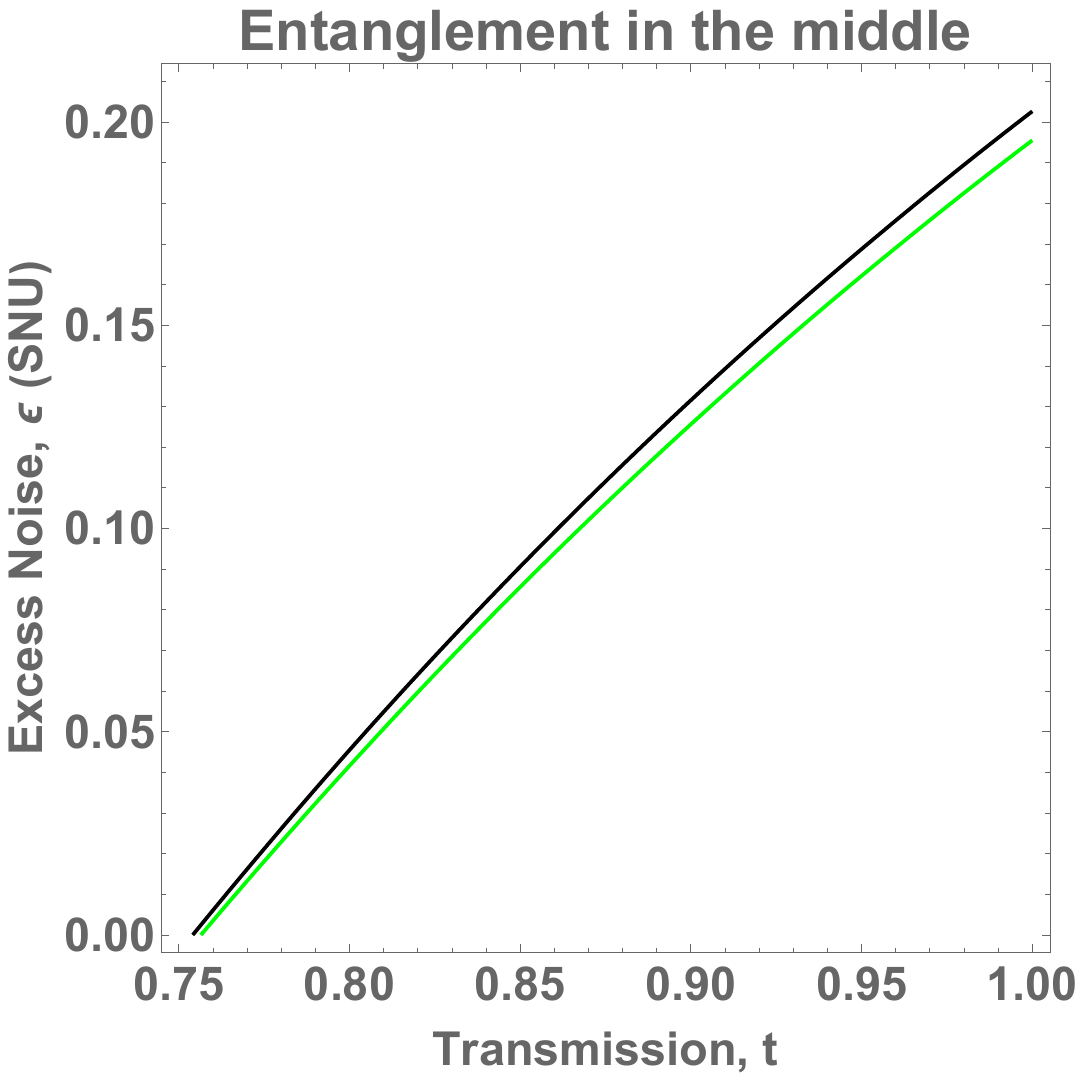}%
        \label{fig:mid_finite}%
    }

    \caption{These plots illustrate the maximum tolerable channel losses and noise in finite size regime with $n=10^7$ measurements, key generation is possible only for channel parameters lying below the plotted curves. The blue curve corresponds to \emph{Distribution 2}, the red to \emph{Distribution 1}, the green to \emph{Distribution 3}, and the black to the GHZ state. With trusted AS noise $V_N = 10 SNU$. The reconciliation efficiency is $\beta = 0.95$, and the variance of the squeezed quadrature is $V = 0.1 SNU$. }
    \label{fig:ckafinite}
\end{figure}

From Fig.~\ref{fig:mid_asy}, we observe that, similar to the case in reverse reconciliation, the protocol with the GHZ state outperforms other protocols. \emph{Distribution 3} closely matches the GHZ state's performance. The adverse influence of trusted noise for \emph{Distribution 4} is noticeable here, though its presence proves beneficial for protocols with \emph{Distribution 3} and the GHZ state.

In contrast to \emph{Distribution 3}, neither the GHZ state nor \emph{Distribution 4} can directly estimate the channel transmission between the dealer and the users. To enable channel parameter estimation, the dealer must occasionally retain, measure, and disclose a subset of the modes. In Fig.~\ref{fig:mid_finite}, we allocate half of the $10^7$ measurements for channel estimation and use the remaining half for key generation. Under these conditions, while the GHZ state still maintains a slight advantage, the performance gap narrows considerably.

\subsection*{Bipartite key post CKA}
Following the generation of a conference key using any of the schemes mentioned in the previous section, bipartite keys can be generated using the conditional data of the remaining users. This data, conditioned on the reference data used for conference key generation, allows for the creation of additional bipartite keys. The following table (Tab.~\ref{tab:bi_scenarios}) summarizes various scenarios and their corresponding key rate expressions.

\begin{table*}[t]
\centering
\renewcommand{\arraystretch}{1.2}
\setlength{\tabcolsep}{4pt}
\caption{Summary of bipartite key rates post QCKA for different scenarios and distributions}
\label{tab:bi_scenarios}
\resizebox{\textwidth}{!}{%
\begin{tabular}{p{0.33\textwidth} c c c}
\toprule
\textbf{Scenario} & \textbf{Tripartite-States Key Rate} & \textbf{Distribution} & \textbf{Key Rate Expression} \\
\midrule

\multirow{2}{*}{Bipartite key rate post CKA with DR}
& $k^{\tilde{B}}(\tilde{B}:\tilde{C}\,|\,A)$
& Distribution 1
& $k^{\tilde{b}_1}(\tilde{b}_1:\tilde{c}_1\,|\,a_1 a_2)$ \\
&
&
& $k^{\tilde{b}_1 \tilde{b}_2}(\tilde{b}_1 \tilde{b}_2:\allowbreak \tilde{c}_1 \tilde{c}_2\,|\,a_1 a_2)$ \\
\midrule

\multirow{2}{*}{Bipartite key rate with DR post CKA with RR}
& $k^{A}(A:\tilde{C}\,|\,\tilde{B})$
& Distribution 1
& $k^{a_1}(a_1:\tilde{c}_1\,|\,b_2 a_2 c_2 \tilde{b}_1)$ \\
&
&
& $k^{a_1 a_2}(a_1 a_2:\allowbreak \tilde{c}_1 \tilde{c}_2\,|\,\tilde{b}_1 \tilde{b}_2)$ \\
\midrule

\multirow{2}{*}{Bipartite key rate with RR post CKA with RR}
& $k^{\tilde{C}}(A:\tilde{C}\,|\,\tilde{B})$
& Distribution 1
& $k^{\tilde{c}_1}(a_1:\tilde{c}_1\,|\,b_2 a_2 c_2 \tilde{b}_1)$ \\
&
&
& $k^{\tilde{c}_1 \tilde{c}_2}(a_1 a_2:\allowbreak \tilde{c}_1 \tilde{c}_2\,|\,\tilde{b}_1 \tilde{b}_2)$ \\
\midrule

\multirow{2}{*}{Bipartite key rate post CKA with Mid}
& $k^{\tilde{B}}(\tilde{B}:\tilde{C}\,|\,\tilde{A})$
& Distribution 3
& $k^{\tilde{b}_1}(\tilde{b}_1:\tilde{c}_1\,|\,a_2 b_2 c_2 \tilde{a}_1)$ \\
&
&
& $k^{\tilde{b}_1 \tilde{b}_2}(\tilde{b}_1 \tilde{b}_2:\allowbreak \tilde{c}_1 \tilde{c}_2\,|\,\tilde{a}_1 \tilde{a}_2)$ \\
\bottomrule
\end{tabular}%
}
\end{table*}

\begin{figure*}[t]
    \centering
    \begin{minipage}[b]{0.24\textwidth}
        \includegraphics[width=\textwidth]{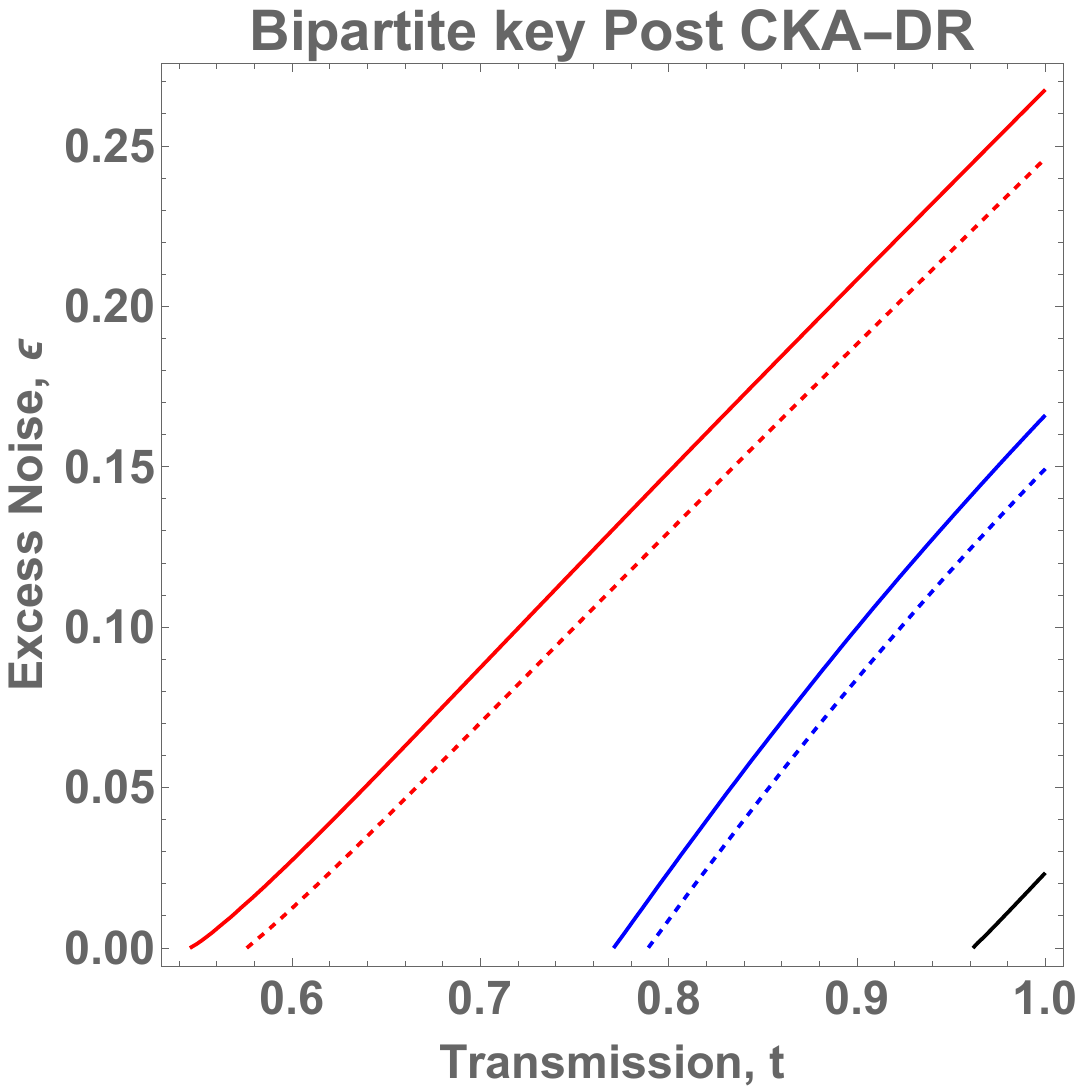}
        \subcaption{}\label{fig:bi_postdr}
    \end{minipage}\hfill
    \begin{minipage}[b]{0.24\textwidth}
        \includegraphics[width=\textwidth]{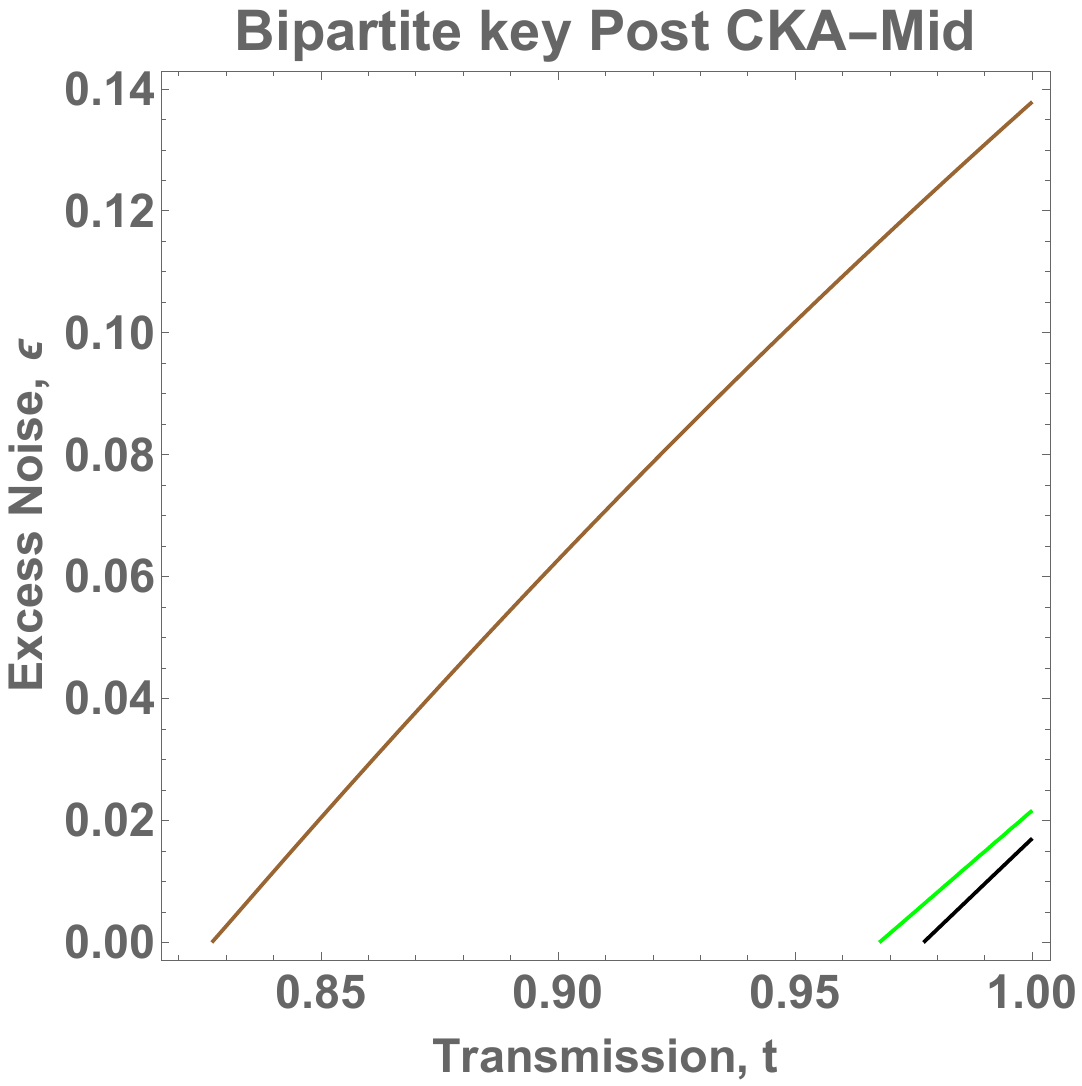}
        \subcaption{}\label{fig:bi_postmid}
    \end{minipage}\hfill
    \begin{minipage}[b]{0.24\textwidth}
        \includegraphics[width=\textwidth]{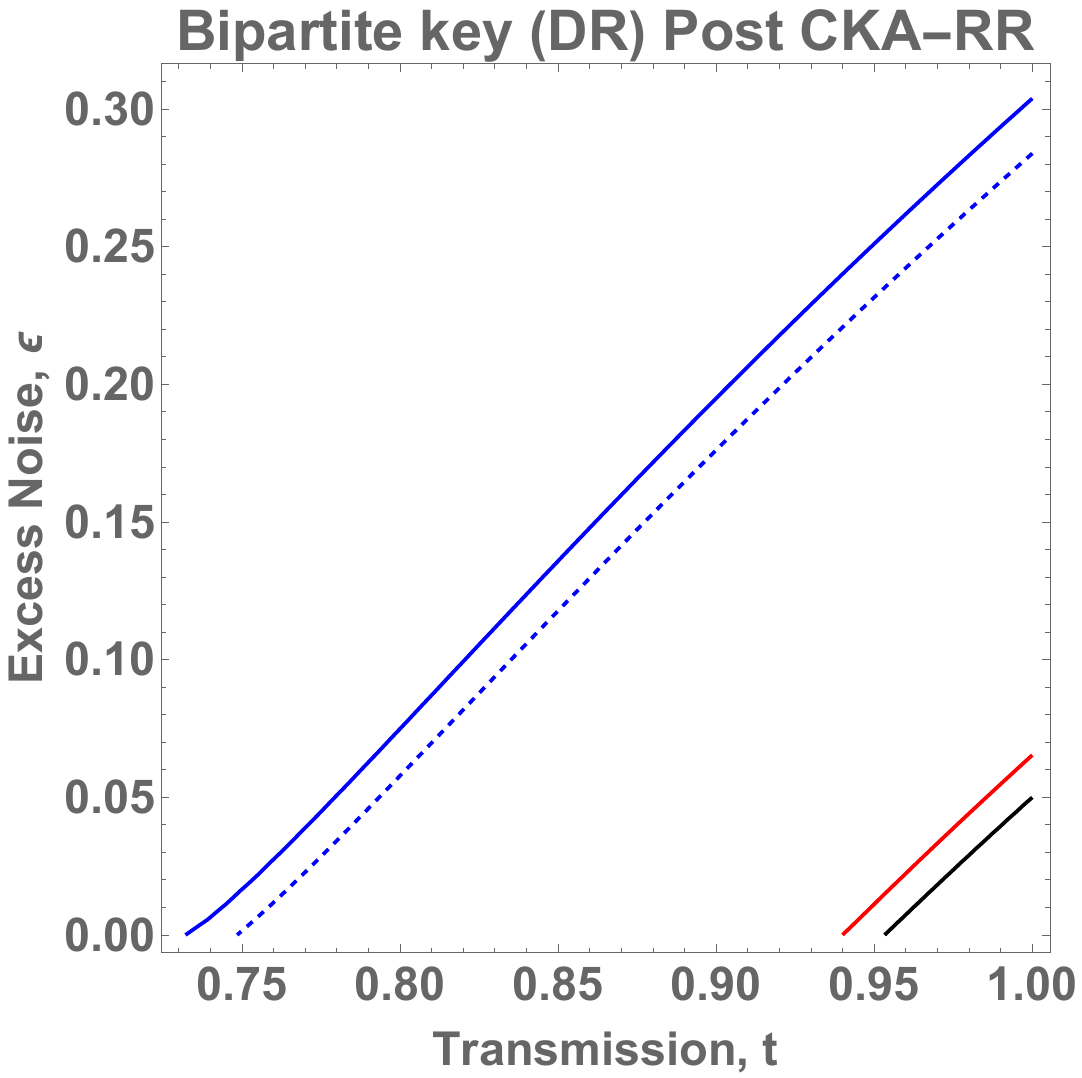}
        \subcaption{}\label{fig:bidr_postrr}
    \end{minipage}\hfill
    \begin{minipage}[b]{0.24\textwidth}
        \includegraphics[width=\textwidth]{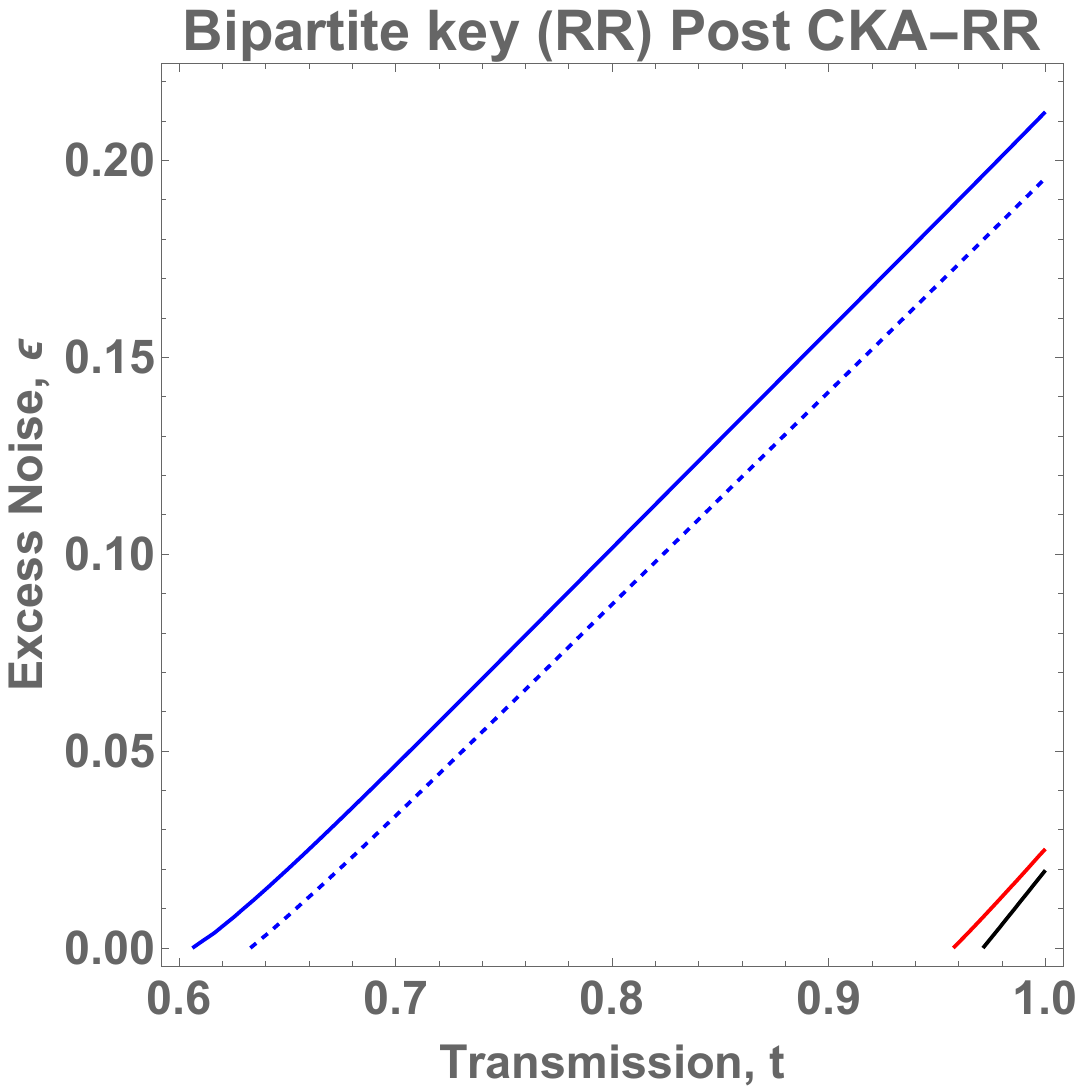}
        \subcaption{}\label{fig:birr_postrr}
    \end{minipage}

    \caption{Maximum tolerable excess noise $\epsilon$ (in SNU) versus transmission $t$ for bipartite keys post CKA generation. Blue: \emph{Distribution 2}; red: \emph{Distribution 1}; green: \emph{Distribution 3}; brown: \emph{Distribution 4}; black: GHZ state. Dashed = finite-size keys, solid = asymptotic. Parameters: $V_N=10$, $\beta=0.95$, squeezed variance $V=0.1$ SNU.}
    \label{fig:bi_postCKA}
\end{figure*}

When comparing the overall multiuser QKD capabilities (CKA and subsequent bipartite key generation) of multipartite states, the six-mode graph state derived from a dual rail cluster state significantly outshines the GHZ state: see Fig.~\ref{fig:bi_postCKA}. For CKA via DR, \emph{Distribution 1} supports CKA and bipartite key generation under a broader range of channel conditions than any other considered multipartite state or strategy. For CKA via RR, \emph{Distribution 2} excels beyond all others. For CKA via Mid, \emph{Distribution 4} facilitates CKA and bipartite keys under more extensive channel conditions.
In the finite-size regime, bipartite keys generated using the GHZ state become infeasible for measurements with fewer than \(n = 10^8\) samples. In contrast, protocols employing dual-rail cluster states enable the generation of bipartite keys after the conference key is established, making them applicable to all the protocols introduced. However, the distribution strategy must be carefully tailored to the specific requirements of the quantum network.

\subsection*{Independent bipartite keys}
In Sec.~\ref{sec:security}, we discussed the utility of analyzing the sum of independent bipartite keys that can be generated by a central user who has higher correlations with other users. This approach is particularly informative in symmetric cases, where the collective key rate helps establish bounds on the symmetric channel parameters required for independent key generation by all the users with the central user. As outlined in Sec.~\ref{classical part}, generating independent keys necessitates one user having a stronger correlation with the rest of the users than between the users themselves.

\begin{figure}[H]
    \centering
    \begin{subfigure}[b]{0.50\columnwidth}
        \includegraphics[width=\columnwidth]{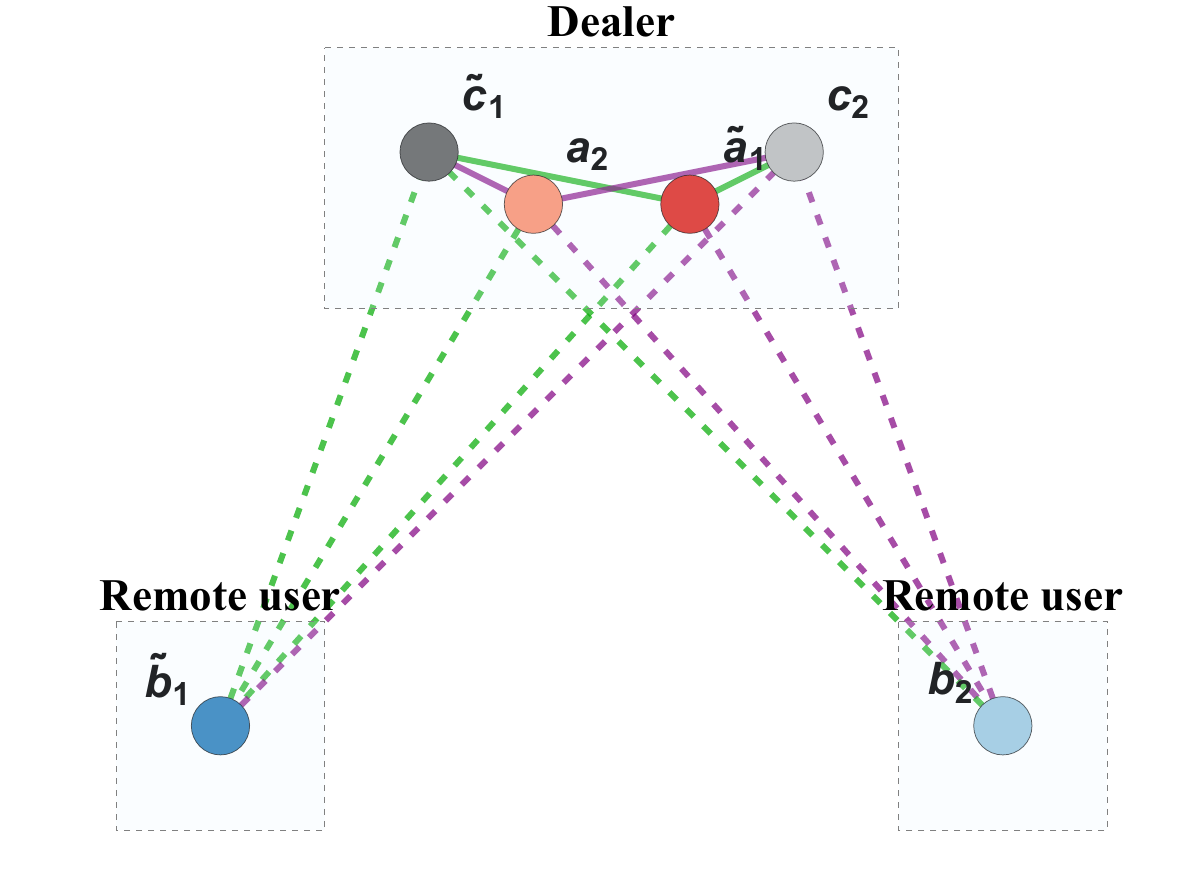}
        \caption{Distribution 5}
        \label{fig:dis_5}
    \end{subfigure}
    \hfill 
    \begin{subfigure}[b]{0.4\columnwidth}
        \includegraphics[width=\columnwidth]{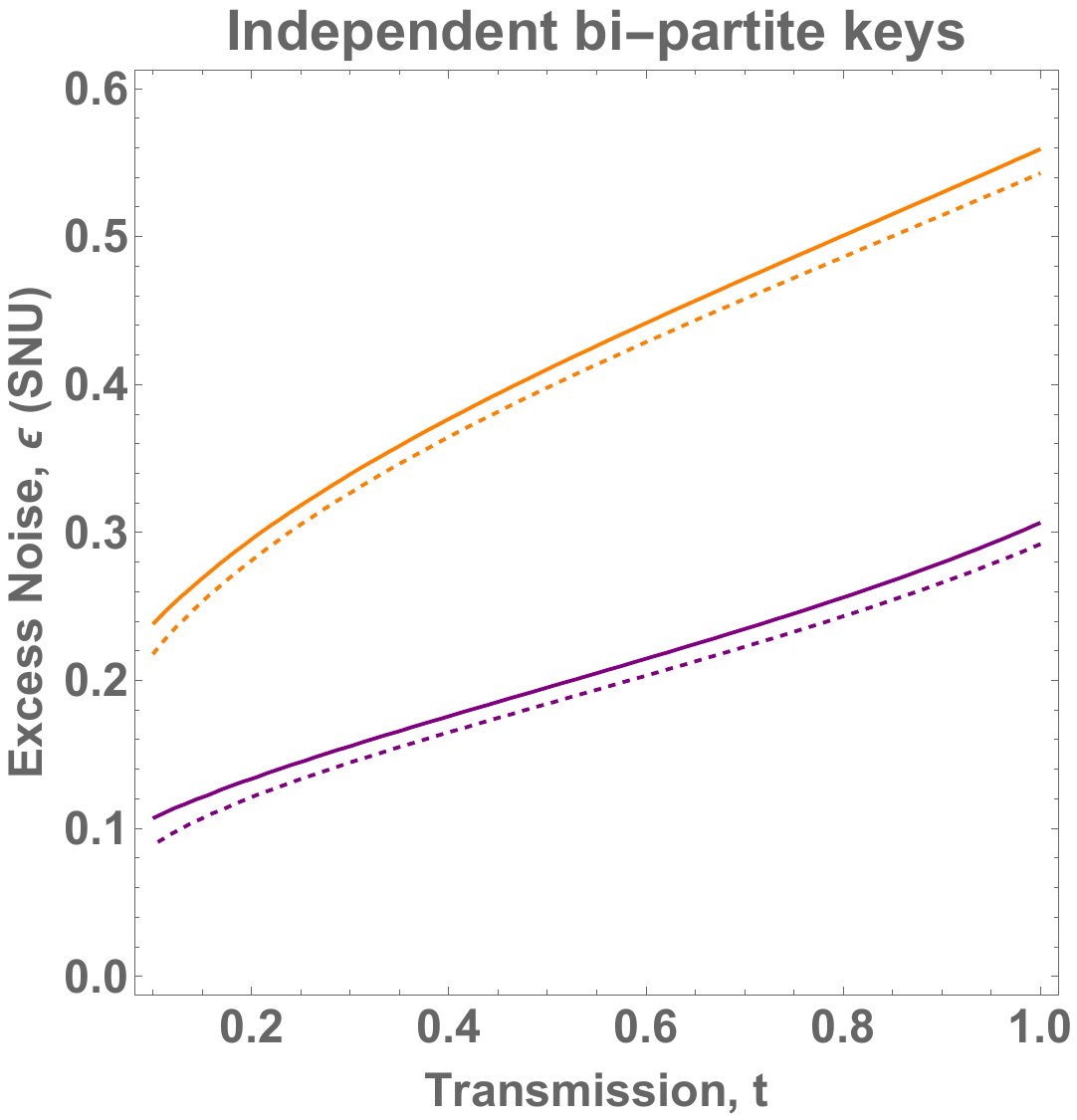}
        \caption{}
        \label{fig:bi_keys}
    \end{subfigure}
\caption{ b) Shows maximum tolerable excess noise $\epsilon$ for varying transmission of the channel $t$ for generating independent bipartite keys between the dealer and the remote users. Orange plots correspond to Distribution 5 (a) and the purple plot corresponds to DAN. Dashed line corresponds to finite size keys ($n=10^7$) and solid lines corresponds to asymptotic keys. For AS noise of  $V_N = 10$, the reconciliation efficiency is $\beta = 0.95$, and the variance of the squeezed quadrature is $V = 0.1$ SNU.}
    \label{fig:fourSubfigures}
\end{figure}
It is more advantageous in terms of classical information processing for these remote users to be uncorrelated. We introduce a \emph{Distribution 5}, depicted in Fig.~\ref{fig:dis_5} , specifically because the remote users are not correlated.  It is evident from Fig.~\ref{fig:bi_keys} that this approach offers a significant enhancement compared to DAN.

\section{Results}\label{sec:results}

In this work, we introduced a novel protocol for three-user conference key generation using dual-rail cluster states, employing the concept of a \textbf{quotient graph state} to create a finite six-mode pure graph state suitable for cryptographic applications. Our comparative analysis with GHZ/W states generated under similar squeezing conditions yielded the following key findings:

\begin{itemize}
    \item \textbf{Asymptotic Regime (CKA):} In the limit of infinite signals, our protocol outperforms GHZ/W states in the direct reconciliation scheme but does not surpass them in reverse reconciliation or entanglement-in-the-middle schemes.

\item \textbf{Finite-Size Regime (CKA):} GHZ states maintain a slight performance advantage for generation of conference key in all schemes explored, direct reconciliation, reverse reconciliation, and entanglement-in-the-middle. However, the performance gap narrows significantly as our protocol leverages multiple modes per user, enhancing channel parameter estimation and mitigating finite-size effects. This improvement brings the key rates closer to those achieved with GHZ states, reducing the impact of statistical fluctuations inherent in finite-size scenarios.

    \item \textbf{Bipartite Key Generation Post-QCKA:} Our protocol significantly outperforms GHZ states in generating bipartite keys after the conference key agreement across all methods (direct reconciliation, reverse reconciliation, and entanglement-in-the-middle). The ability to derive additional bipartite keys post-QCKA enhances the overall efficiency and security of the quantum network.

    \item \textbf{Bipartite Key Generation Without QCKA:} When compared to downstream access networks utilizing two-mode squeezed vacuum states with equivalent squeezing, our protocol achieves superior performance in generating bipartite keys without involving QCKA. This demonstrates the versatility of our protocol in supporting both conference and individual key generation within the network.

    \item \textbf{Robustness to Imperfections:} The protocol maintains its advantages even when using impure squeezed states (provided the impurity is trusted), demonstrating robustness against experimental imperfections. This robustness is critical for practical implementations where perfect state preparation is challenging.
\end{itemize}

\begin{table*}[t]
\centering
\renewcommand{\arraystretch}{1.2}
\setlength{\tabcolsep}{4pt}
\caption{Recommended distribution strategies and corresponding information reconciliation methods for multi-user QKD}
\label{tab:dist_strategies}
\resizebox{\textwidth}{!}{%
\begin{tabular}{|c|p{0.25\textwidth}|p{0.25\textwidth}|p{0.25\textwidth}|}
\hline
\textbf{Strategy} & \textbf{Operational Context} & \textbf{Information Reconciliation} & \textbf{Advantages} \\
\hline
\textbf{Distribution 1} & Dealer is also a user & Direct Reconciliation, Reverse Reconciliation & Optimal for maximal distance; robust against trusted noise. \\
\hline
\textbf{Distribution 2} & Dealer is also a user & Direct Reconciliation, Reverse Reconciliation & Better performance in finite-size regime; enhanced channel parameter estimation via multiple correlations. \\
\hline
\textbf{Distribution 3} & The dealer distributes three modes. Measures and publicly discloses the remaining three. & Entanglement-in-the-Middle & Suitable for decentralized conference key agreement. \\
\hline
\textbf{Distribution 4} & Fully distributed; no dealer involvement after state distribution. & Entanglement-in-the-Middle & Enables fully peer-to-peer operation. \\
\hline
\textbf{Distribution 5} & Dealer sends two uncorrelated modes to remote users & Direct Reconciliation (CKA), Reverse Reconciliation & Best suited for generating separate keys between dealer and users. \\
\hline
\end{tabular}%
}
\end{table*}

\section{Conclusion}\label{sec:conclusion}

In this work, we have introduced a novel protocol for three-user conference key generation using dual-rail cluster states, employing the concept of a \textit{quotient graph state} to create a six-mode pure graph state suitable for cryptographic applications. Our comparative analysis with GHZ/W states generated under similar squeezing conditions has yielded significant insights.

In the asymptotic regime, where the number of signals approaches infinity, our protocol outperforms GHZ/W states in the direct reconciliation scheme but does not surpass them in reverse reconciliation or entanglement-in-the-middle schemes. In the finite-size regime, GHZ states still perform marginally better in each scheme. However, the performance gap narrows significantly as our protocol benefits from each user accessing multiple modes, which enhances channel parameter estimation and mitigates finite-size effects. This improved estimation reduces the impact of statistical fluctuations, bringing the maximum achievable distances closer to those achieved with GHZ states.

Additionally, our protocol significantly outperforms GHZ/W states in generating bipartite keys after the conference key agreement across all methods. It also surpasses downstream access networks utilizing two-mode squeezed vacuum states in generating bipartite keys without involving QCKA. This versatility demonstrates the adaptability of our protocol in supporting both conference and individual secure communications within the network.

Our protocol maintains its advantages even when using impure squeezed states, showcasing robustness against experimental imperfections. This robustness is critical for practical implementations where perfect state preparation is challenging.

These results underscore the importance of tailoring quantum cryptographic solutions to the specific demands of quantum networks. The use of dual-rail cluster states and quotient graph states provides a pathway to more robust, flexible, and efficient quantum cryptographic protocols with continuous-variable systems.

Looking ahead, the insights gained from this work highlight several promising directions for future research. Exploring alternative multipartite states with configurations that allow users to access multiple modes could further enhance key rates and robustness in quantum communication protocols. Extending the protocol to networks with more than three users and examining the performance benefits in larger, more complex quantum networks is a natural progression. Developing advanced methods for channel parameter estimation that leverage access to multiple modes per user could further mitigate finite-size effects and improve the security and efficiency of the protocols. A crucial next step for this protocol is its experimental implementation, which presents a set of engineering challenges. For scalable, time-multiplexed resources, a central challenge is the engineering of high-speed optical switches to demultiplex and distribute modes to multiple users. While minimizing loss is ideal, the security framework can accommodate these losses if the device is trusted and well-characterized, shifting the problem towards precise calibration. Additionally, extending the security analysis to consider more sophisticated eavesdropping strategies would verify the protocol's robustness against a wider range of attacks.

\newpage

\section*{Data availability}
The data that support the findings of this study are available from the corresponding author upon reasonable request.

\section*{Acknowledgements}
I would like to acknowledge Vladyslav Usenko, Ivan Derkach, Tobias Gehring, Christoph Pacher, Florian Kanitschar, and Ulrik Lund Andersen for their valuable discussions. A. O. acknowledges the project 8C22002 (CVStar) of the Czech Ministry of
Education, Youth and Sports, which has received funding from the
European Union's Horizon 2020 research and innovation framework
programme under grant agreement No. 731473 and 101017733, and the
project No. 21-44815L of the Czech Science Foundation. A. O. also
acknowledges the project IGA-PrF-2025-010 of Palacký University Olomouc.

\appendix

\section{Dual rail cluster state to six mode graph state}\label{sec:dual to six}

In Sec.~\ref{sec:dual}, we introduced a method to generate a pure six-mode graph state from a dual-rail cluster state. Here, we provide a justification for this method. Since we are working with Gaussian states, it suffices to show that, after appropriately grouping the nodes as suggested by the coloring scheme in Fig.~\ref{fig:dual}, the covariance matrix of the resulting state corresponds to that of a pure six-mode graph state.

The dual-rail cluster state, denoted here by $\rho_{\text{dual}}$, consists of two parallel chains (rails) of quantum states: Rail~$A$ and Rail~$B$. Each node (quantum state) in these rails is uniquely identified by its rail ($A$ or $B$) and its position $i$ along that rail as illustrated in Fig.~\ref{fig:dualrail_cluster}.
\begin{figure*}[t]
  \centering
  \begin{tikzpicture}[scale=2] 
    \node (v0) at (6.0,0.07) {};
    \node (v1) at (6.0,0.45) {};
    \node (v2) at (5.0,0.45) {$\rho^{(A)}_1$};
    \node (v3) at (5.0,0.07) {$\rho^{(B)}_1$};
    \node (v4) at (4.0,0.07) {$\rho^{(B)}_2$};
    \node (v5) at (4.0,0.45) {$\rho^{(A)}_2$};
    \node (v6) at (3.0,0.07) {};
    \node (v7) at (3.0,0.45) {};
    \node (v12) at (2.0,0.07) {$\rho^{(B)}_{(3n-1)}$};
    \node (v13) at (2.0,0.45) {$\rho^{(A)}_{(3n-1)}$};
    \node (v14) at (1.0,0.45) {$\rho^{(A)}_{3n}$};
    \node (v15) at (1.0,0.07) {$\rho^{(B)}_{3n}$};
    \node (v16) at (0.0,0.07) {};
    \node (v17) at (0.0,0.45) {};
    \draw[dashed, line width=1pt, color=purple] (v0) -- (v2);
    \draw[dashed, line width=1pt, color=purple] (v0) -- (v3);
    \draw[dashed, line width=1pt, color=green] (v1) -- (v2);
    \draw[dashed, line width=1pt, color=green] (v1) -- (v3);
    \draw[color=green] (v2) -- (v4);
    \draw[color=green] (v2) -- (v5);
    \draw[color=purple] (v3) -- (v4);
    \draw[color=purple] (v3) -- (v5);
    \draw[dashed, line width=1pt, color=purple] (v4) -- (v6);
    \draw[dashed, line width=1pt, color=purple] (v4) -- (v7);
    \draw[dashed, line width=1pt, color=green] (v5) -- (v6);
    \draw[dashed, line width=1pt, color=green] (v5) -- (v7);
    \draw[decorate, decoration={snake, amplitude=0.5mm, segment length=3mm}] (v6) -- (v7);
    \draw[dashed, line width=1pt, color=purple] (v6) -- (v12);
    \draw[dashed, line width=1pt, color=purple] (v6) -- (v13);
    \draw[dashed, line width=1pt, color=green] (v7) -- (v12);
    \draw[dashed, line width=1pt, color=green] (v7) -- (v13);
    \draw[color=purple] (v12) -- (v14);
    \draw[color=purple] (v12) -- (v15);
    \draw[color=green] (v13) -- (v14);
    \draw[color=green] (v13) -- (v15);
    \draw[dashed, line width=1pt, color=green] (v14) -- (v16);
    \draw[dashed, line width=1pt, color=purple] (v15) -- (v16);
    \draw[dashed, line width=1pt, color=green] (v14) -- (v17);
    \draw[dashed, line width=1pt, color=purple] (v15) -- (v17);
  \end{tikzpicture}
  \caption{Dual-rail cluster state $\rho_{\text{dual}}$ comprising two parallel rails ($A$ and $B$). Each node is labelled by its rail and position $i$.}
  \label{fig:dualrail_cluster}
\end{figure*}

To construct the six-mode graph state, we group the nodes from each rail into three groups by selecting every third node. For each node at position $i$ in rail $R$, we define the constituent state $\rho^{(R)}_{i}$ as the reduced density operator obtained by tracing out all other nodes from the dual-rail cluster state $\rho_{\text{dual}}$:
\[
\rho^{(R)}_{i} = \operatorname{Tr}^{(R)}_{i} \left( \rho_{\text{dual}} \right),
\]
where $\operatorname{Tr}^{(R)}_{i}$ denotes the partial trace over all modes except the one at position $i$ in rail $R$.
We then form statistical mixtures by averaging the constituent states within each group. For each group \( j \in \{1, 2, 3\} \) in rail \( R \), the mixed state \( \boldsymbol{\rho}^{(R)}_j \) is defined as:
\begin{equation}\label{eq:group_mixed_state}
    \boldsymbol{\rho}^{(R)}_j = \frac{1}{n} \sum_{m=1}^{n} \rho^{(R)}_{3m - (3 - j)},
\end{equation}
where \( n \) is the total number of grouping repetitions along each rail. Specifically, the positions \( i \) of the nodes included in each group are:
\begin{itemize}
    \item For \( j = 1 \): positions \( i = 3m - 2 \),
    \item For \( j = 2 \): positions \( i = 3m - 1 \),
    \item For \( j = 3 \): positions \( i = 3m \),
\end{itemize}
with \( m \in \{1, 2, \ldots, n\} \).

By forming these statistical mixtures, we create six mixed states in total (three groups in each of the two rails). The quadrature operators associated with these mixed states are denoted \( \boldsymbol{x}^{(R)}_j \) and \( \boldsymbol{p}^{(R)}_j \), where \( j \in \{1, 2, 3\} \) and \( R \in \{ A, B \} \).

Since the constituent states \( \rho^{(R)}_{i} \) are uncorrelated in the statistical mixture, the variances and covariances of the quadratures of these mixed states are given by:
\begin{equation}\label{eq:variance_mixed_state}
    \left\langle \left( \boldsymbol{X}^{(R)}_j \right)^2 \right\rangle = \frac{1}{n} \sum_{m=1}^{n} \left\langle \left( X^{(R)}_{3m - (3 - j)} \right)^2 \right\rangle,
\end{equation}
\begin{equation}\label{eq:covariance_mixed_states}
    \frac{1}{2} \left\langle [ \boldsymbol{X}^{(R)}_p , \boldsymbol{X}^{(S)}_q ]_+ \right\rangle = \frac{1}{2n} \sum_{m=1}^{n} \left\langle [X^{(R)}_{3m - (3 - p)} , X^{(S)}_{3m - (3 - q)} ]_+ \right\rangle,
\end{equation}
where \( R, S \in \{ A, B \} \), \( p = 2, \) and \( q \in \{1, 3\} \), \( X^{(R)}_{i} \in \{ x^{(R)}_{i}, p^{(R)}_{i} \} \), with \([.]_+\) denoting the anti-commutator.
\begin{equation}
    \frac{1}{2}\left\langle[ \boldsymbol{X}^{(k)}_1 ,\boldsymbol{X}^{(l)}_3 ]_+\right\rangle=\frac{1}{2n} \sum_{m=2}^{n} \left\langle [X^{(k)}_{3m-3} ,X^{(l)}_{3m-2} ]_+\right\rangle
\end{equation}
The covariance matrix of the six-mode graph state is then constructed from these variances and covariances. We arrange the quadrature operators in the following order:
\begin{equation}\label{eq:quadrature_vector}
\begin{aligned}
\mathbf{X} = \bigl(&\boldsymbol{x}^{(A)}_1, \boldsymbol{p}^{(A)}_1, \boldsymbol{x}^{(B)}_1, \boldsymbol{p}^{(B)}_1, \boldsymbol{x}^{(A)}_2, \boldsymbol{p}^{(A)}_2, \\
&\boldsymbol{x}^{(B)}_2, \boldsymbol{p}^{(B)}_2, \boldsymbol{x}^{(A)}_3, \boldsymbol{p}^{(A)}_3, \boldsymbol{x}^{(B)}_3, \boldsymbol{p}^{(B)}_3 \bigr)^{\!T}.
\end{aligned}
\end{equation}

With this ordering, the covariance matrix \( \boldsymbol{\sigma} \) of the six-mode graph state is given by:
\begin{equation}\label{eq:six_mode_covariance_matrix}
\boldsymbol{\sigma} =
\begin{pmatrix}
   \boldsymbol{V} & \boldsymbol{0} & \boldsymbol{C} & -\boldsymbol{C} & \frac{n-1}{n} \boldsymbol{C} & \frac{n-1}{n} \boldsymbol{C} \\
   \boldsymbol{0} & \boldsymbol{V} & \boldsymbol{C} & -\boldsymbol{C} & -\frac{n-1}{n} \boldsymbol{C} & -\frac{n-1}{n} \boldsymbol{C} \\
   \boldsymbol{C} & \boldsymbol{C} & \boldsymbol{V} & \boldsymbol{0} & \boldsymbol{C} & -\boldsymbol{C} \\
   -\boldsymbol{C} & -\boldsymbol{C} & \boldsymbol{0} & \boldsymbol{V} & \boldsymbol{C} & -\boldsymbol{C} \\
   \frac{n-1}{n} \boldsymbol{C} & -\frac{n-1}{n} \boldsymbol{C} & \boldsymbol{C} & \boldsymbol{C} & \boldsymbol{V} & \boldsymbol{0} \\
   \frac{n-1}{n} \boldsymbol{C} & -\frac{n-1}{n} \boldsymbol{C} & -\boldsymbol{C} & -\boldsymbol{C} & \boldsymbol{0} & \boldsymbol{V}
\end{pmatrix}.
\end{equation}
Here, \( \boldsymbol{V} \) and \( \boldsymbol{C} \) are \( 2 \times 2 \) matrices representing the variances and covariances of the quadrature operators, respectively. These can be expressed as a function related to the squeezed quadrature variance $v$ and excess AS noise of the squeezed state employed in generating the dual rail cluster state Eq.~\ref{eq:covariance2}. Specifically, \( \boldsymbol{V} \) is the covariance matrix of a single mixed state, and \( \boldsymbol{C} \) captures the correlations between different mixed states.

The factor \( \frac{n-1}{n} \) within the covariance matrix is due to the fact that the count of vertices in a group that includes the extreme vertices cannot match the number of edges linking them. This factor makes the multipartite Gaussian state impure.   One can purify this impurity, introduced due to finite states, by introducing additional modes. Alternatively, one can consider a pure six mode covariance matrix by removing the factor \( \frac{n-1}{n} \). Introducing additional modes to purify the covariance matrix Eq.~\ref{eq:six_mode_covariance_matrix} would give us a tighter bound than considering the covariance matrix Eq.~\ref{eq:6modecm}. That is, considering the pure six mode covariance matrix to estimate the Holevo bound is more pessimistic than considering a purified state corresponding to the covariance matrix Eq.~\ref{eq:six_mode_covariance_matrix} to estimate the Holevo bound.  For sufficiently large $n$, the covariance matrix $\boldsymbol{\sigma}$ can be approximated to Eq.~\ref{eq:6modecm} a pure six mode state, and hence the improvement achieved through purification is insignificant.

\section{Congruence Transformation for Covariance Matrices}\label{sec:transformation}

Given an original covariance matrix \( \boldsymbol{\Sigma} \) of dimensions \( 12n \times 12n \), of the modes of the dual-rail cluster state under consideration for grouping, we can derive the covariance matrix \( \boldsymbol{\sigma} \) of dimensions \( 12 \times 12 \) corresponding to the six-mode graph state. This is achieved by applying a congruence transformation using a matrix \( \mathbf{A} \) of size \( 12 \times 12n \), such that:
\begin{equation}\label{eq:covariance_transformation}
    \boldsymbol{\sigma} = \frac{1}{n}\mathbf{A} \boldsymbol{\Sigma} \mathbf{A}^\intercal.
\end{equation}
The ordering of quadratures in covariance matrix of the modes of dual rail cluster states considered for grouping is:
\begin{equation}
    (x^{(a)}_1,p^{(a)}_1,x^{(b)}_1,p^{(b)}_1,x^{(a)}_2,p^{(a)}_2,x^{(b)}_2,p^{(b)}_2......)
\end{equation}
Here, \( \mathbf{A}^\intercal \) denotes the transpose of \( \mathbf{A} \). This operation effectively projects the original covariance matrix onto a lower-dimensional subspace defined by \( \mathbf{A} \), capturing the relevant quadrature correlations between the groups (color classes) of modes.

\( \mathbf{A} \) consists of vectors that indicate the modes present within a color class.
\begin{equation}\label{eq: Aij}
    \mathbf{A}_{ij} = \begin{cases}
1 & \text{if } j \equiv i \pmod{11} \\
0 & \text{otherwise}
\end{cases}
\end{equation}

A congruence transformation for the covariance matrix is possible provided the modes within each color class remain uncorrelated. However, if correlations exist among modes within a color class, the congruence transformation outlined above does not yield the covariance matrix of the statistical mixture defined by such coloring scheme.

\begin{center}
\textbf{Note}: \(\mathbf{X}_{\text{six-mode}} \neq \frac{1}{\sqrt{n}} \mathbf{A} \mathbf{X}_{\text{dual}}\)
\end{center}

Based on the above congruence transformation, one might be inclined to define the quadrature operation of the six-mode graph state as a linear combination of the quadrature operators from the constituent states in the dual rail cluster state. However, it is essential to recognize that statistical mixtures cannot be represented by linear transformations of quadrature operators because the modes are not coherently combined. Instead, the covariance matrix of the mixture is the average of the covariance matrices of the constituent states.

\subsection{Congruence  transformation for adjacency matrices}
A complex-weighted adjacency matrix is used in continuous-variable quantum computing to represent Gaussian pure states, extending the real-weighted graphs of ideal cluster states to more realistic, physical states \cite{PhysRevA.83.042335}. This formalism simplifies the analysis of entanglement structures, Gaussian operations, and deviations from ideal behaviors, which are essential for applications in CV quantum computing. Thus, we define a similar transformation to the one used in the previous section, enabling us to derive the adjacency matrix for a six-mode graph state $\boldsymbol{Z}_{six-mode}$ from the adjacency matrix of a dual-rail cluster state $\boldsymbol{Z}_{dual}$.
\begin{equation}
    \boldsymbol{Z}_{six-mode}=\frac{1}{n}\boldsymbol{B}\boldsymbol{Z}_{dual}\boldsymbol{B^T}
\end{equation}

\begin{equation}\label{eq: Bij}
    \mathbf{B}_{ij} = \begin{cases}
1 & \text{if } j \equiv i \pmod{5} \\
0 & \text{otherwise}
\end{cases}
\end{equation}
\begin{figure}[H]
    \centering

    \begin{subfigure}{\columnwidth}
        \centering
        \includegraphics[width=\columnwidth]{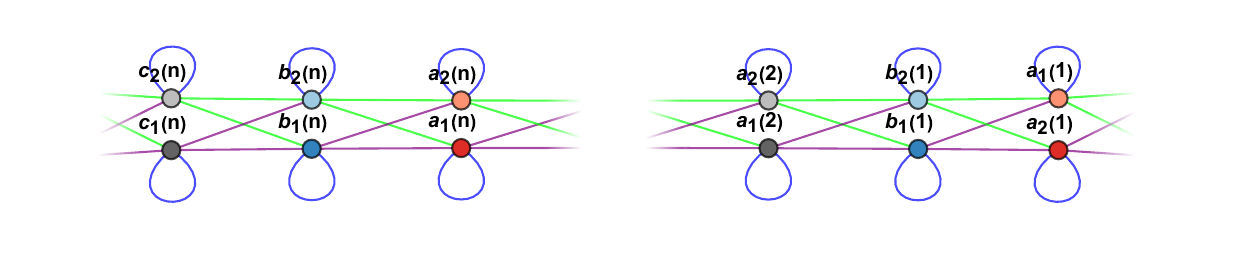} 
        \caption{Complex weighted graph of dual rail cluster state}
        \label{fig:sub1}
    \end{subfigure}

    \vspace{0.5cm} 

    \begin{subfigure}{\columnwidth}
        \centering
        \includegraphics[width=0.8\columnwidth]{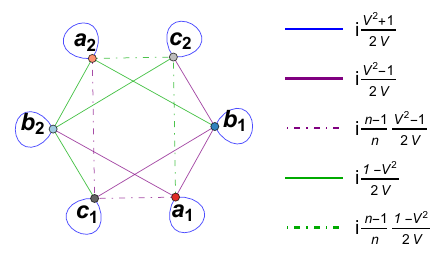} 
        \caption{Complex weighted graph of quotient graph derived from dual rail cluster state.}
        \label{fig:sub2}
    \end{subfigure}

    \caption{}
    \label{fig:main}
\end{figure}
\section{Rationale for the coloring strategy}

To convert the infinitely repeating dual-rail cluster lattice into a finite resource that is easy to manipulate in security proofs, we map groups of physical modes onto \emph{color classes} and then merge every class into a single vertex; the resulting state is called a \emph{quotient graph state}.  The coloring we adopt is engineered to satisfy two operational requirements:
\begin{enumerate}[label=(\roman*)]
\item \textbf{Neighbourhood preservation} --- The quotient must retain all edges that enter the closed neighbourhood $N[v]$ of at least one vertex $v$ in the original lattice, so that any pair of modes that were immediate neighbours remain directly correlated after the homomorphism.
\item \textbf{Purity} ---The quotient state should be a \emph{pure} Gaussian graph state; purity simplifies both covariance-matrix algebra and the security analysis in Sec.~\ref{sec:security}.
\end{enumerate}
The procedure is generic and can be reused for other cluster topologies; we describe it once here and apply it to the dual-rail lattice below.

\paragraph{Step~1: Identify a unit cell that encloses a full closed neighbourhood.} Let $G=(V,E)$ be the graph of the open-ended lattice.  We search for the smallest \emph{induced} subgraph $G_{\text{cell}}\subset G$ that contains a vertex $v$ together with all vertices in $N[v]$.  For the dual-rail lattice, this unit cell contains six vertices: two from each rail at positions $3m-2$, $3m-1$, and $3m$ ($m\in\mathbb{Z}$).

\paragraph{Step~2: Assign one color per vertex inside the cell.} Each vertex of $G_{\text{cell}}$ receives a distinct color $c\in{1,\dots,|V(G_{\text{cell}})|}$.  Translational tiling of the cell propagates this coloring across the infinite lattice, producing exactly six global color classes in the dual-rail case.

\paragraph{Step3: Form the quotient graph.} Vertices that share the same color are identified (fused) into a single vertex.  If colors $c_i$ and $c_j$ appeared on adjacent physical vertices, the corresponding quotient vertices are connected by an edge.  Algebraically, if $Z_{\text{dual}}$ is the complex-weighted adjacency matrix of the dual-rail lattice, the quotient adjacency matrix is obtained via the congruence transformation.
\begin{equation}
Z_{\text{quot}} = \frac{1}{n}BZ_{\text{dual}}B^{\mathsf T},
\end{equation}
where the selector matrix $B$ is defined in Sec.~\ref{sec:transformation} and $n$ is the number of cell repetitions.

\paragraph{Step 4: Verify purity and neighbourhood fidelity.} The same transformation applied to the lattice covariance matrix $\Sigma$ yields
\begin{equation}
\sigma = \frac{1}{n}A\Sigma A^{\mathsf T},
\end{equation}
with $A$ given in Eq.~\ref{eq: Aij}.  Taking the limit $n\to\infty$ removes boundary terms proportional to $(n-1)/n$, leaving the six-mode covariance matrix of Eq.~\ref{eq:6modecm} which is pure ($Det[\sigma]=1$).  Moreover, every edge that linked a vertex to its first neighbours inside $G_{\text{cell}}$ survives in $Z_{\text{quot}}$, so immediate–neighbour correlations are identical to those of the parent lattice.

\textbf{Result for the dual-rail lattice.} Applying the above steps to the dual-rail cluster yields the coloring shown in Fig.~\ref{fig:dual}.  The six color classes correspond to the vertices ${a_1,a_2,b_1,b_2,c_1,c_2}$ of Fig.~\ref{fig:6graph}.  This six-mode pure quotient graph state is the smallest resource that simultaneously preserves all closed–neighbourhood correlations and admits a compact security proof.

\bigskip
\noindent\textbf{Reuse for other lattices.} For any future cluster geometry one simply (i) locates a cell containing a full neighbourhood, (ii) assigns unique colors to its vertices, and (iii) repeats the tiling.  The size of the quotient graph is bounded below by the cardinality of that neighbourhood and above by the chromatic number of the lattice.
\section{Parameter estimation}\label{sec:PE}

To estimate the channel transmission \( t \), we use the fact that the correlation between the measurements of the dealer and the remote user scales as \( \sqrt{t} \). Specifically, we can estimate the channel transmission from the following relation:

\begin{equation}
\sqrt{t}\, \mathcal{C} = \frac{1}{n} \sum_{i=1}^n d_i u_i := C_{du},
\end{equation}

where \( \mathcal{C} = (1/2)\langle [ X^d, X^u ]_+ \rangle \) is the expectation value of the anticommutator of the quadrature operators \( X^d \) and \( X^u \), measured by the dealer and the user before the state is sent through the channel. Here, \( d_i \) and \( u_i \) are the measurement outcomes corresponding to these quadratures for the dealer and the remote user, respectively.

We define \( C_{du} \) as:

\begin{equation}
C_{du} := \frac{1}{n} \sum_{i=1}^n d_i u_i,
\end{equation}

which serves as an estimator for \( \sqrt{t}\, \mathcal{C} \).

Rearranging the equation, we express the transmission \( t \) as:

\begin{equation}
t = \left( \frac{C_{du}}{\mathcal{C}} \right)^2.
\end{equation}

Since \( C_{du} \) is a random variable due to measurement fluctuations, we are interested in the variance of our estimator for \( t \):

\begin{equation}
\operatorname{Var}[t] = \frac{\operatorname{Var}[ C_{du}^2 ]}{\mathcal{C}^4}.
\end{equation}

Assuming that \( C_{du} \) is approximately normally distributed (by the Central Limit Theorem, since it is the average of \( n \) independent measurements), \( C_{du}^2 \) follows a noncentral chi-squared distribution with one degree of freedom. The variance of \( C_{du}^2 \) is given by:

\begin{equation}
\operatorname{Var}[ C_{du}^2 ] = 2 \left( \operatorname{Var}[ C_{du} ] \right)^2 + 4 \left( \mathbb{E}[ C_{du} ] \right)^2 \operatorname{Var}[ C_{du} ].
\end{equation}

Since \( C_{du} \) is an average over \( n \) terms, its variance is:

\begin{equation}
\operatorname{Var}[ C_{du} ] = \frac{1}{n} \operatorname{Var}[ X^d \tilde{X}^u ],
\end{equation}

where \( \tilde{X}^u \) is the quadrature operator of the state that the user possesses after transmission through the channel, and \( X^u \) is the quadrature operator of the state before it is sent through the channel.

Substituting back, we have:

\begin{equation}
\operatorname{Var}[t] = \frac{2 \left( \operatorname{Var}[ C_{du} ] \right)^2 + 4 \left( \mathbb{E}[ C_{du} ] \right)^2 \operatorname{Var}[ C_{du} ] }{ \mathcal{C}^4 }.
\end{equation}

For large \( n \), the term \( 2 \left( \operatorname{Var}[ C_{du} ] \right)^2 \) is of order \( 1/n^2 \) and can be neglected compared to \( 4 \left( \mathbb{E}[ C_{du} ] \right)^2 \operatorname{Var}[ C_{du} ] \), which is of order \( 1/n \). Therefore, we approximate:

\begin{equation}
\operatorname{Var}[t] \approx \frac{4 \left( \mathbb{E}[ C_{du} ] \right)^2 \operatorname{Var}[ C_{du} ] }{ \mathcal{C}^4 }.
\end{equation}

Substituting \( \operatorname{Var}[ C_{du} ] = \frac{1}{n} \operatorname{Var}[ X^d \tilde{X}^u ] \) and \( \mathbb{E}[ C_{du} ] = \mathbb{E}[ X^d \tilde{X}^u ] \), we obtain:

\begin{equation}
\operatorname{Var}[t] \approx \frac{4 \left( \mathbb{E}[ X^d \tilde{X}^u ] \right)^2 \operatorname{Var}[ X^d \tilde{X}^u ] }{ n \mathcal{C}^4 }=\frac{4 t \operatorname{Var}[ X^d \tilde{X}^u ] }{ n \mathcal{C}^2 }.
\end{equation}

In \cite{oruganti2024continuousvariablequantumkeydistribution}, we have shown that combining measurement outcomes of both quadratures can increase the precision of our estimation of the channel transmission. Here, we apply the same principle by leveraging multiple correlations to decrease the variance of our estimators. For example, when the dealer has two modes that are correlated with the user's measurements, we can define the estimator:

\begin{equation}
\sqrt{t} \left( \frac{ \mathcal{C}^a + \mathcal{C}^b }{2} \right) = \frac{1}{2n} \left( \sum_{i=1}^n d_i^a u_i + \sum_{i=1}^n d_i^b u_i \right) := C_{du},
\end{equation}

where \( \mathcal{C}^i = \langle [ X_i^d, X^u ]_+ \rangle \) for \( i = a, b \), and \( X_i^d \) are the quadratures of the modes possessed by the dealer.

For a symmetric case where \( \mathcal{C}^a = \mathcal{C}^b \) and \( \operatorname{Var}[ X_a^d \tilde{X}^u ] = \operatorname{Var}[ X_b^d \tilde{X}^u ] \), the new variance of the estimator for \( t \) becomes:

\begin{equation}
\operatorname{Var}[t] \approx \frac{4 t \operatorname{Var}[ X^d \tilde{X}^u ] }{ 2 n \mathcal{C}^2 }.
\end{equation}

We can generalize this result for any number \( k \) of symmetric correlations used to estimate the channel transmission:

\begin{equation}
\operatorname{Var}[t] \approx \frac{4 t \operatorname{Var}[ X^d \tilde{X}^u ] }{ k n \mathcal{C}^2 }.
\end{equation}

Thus, by leveraging multiple correlated modes, we reduce the variance of the estimator for the transmission \( t \) by a factor of \( k \), enhancing the precision of our estimation.

We estimate the excess noise as follows.
\begin{equation}
    V_{\epsilon}=\frac{1}{n}\sum_i^n u^2_i+t (1-Var[X^u])-1
\end{equation}
The variance of the excess noise estimator can be approximated to.
\begin{equation}
     Var[V_{\epsilon}]\approx\frac{2}{n}(Var[\tilde{X}^u])^2+ Var[t] (1-Var[X^u])^2
\end{equation}

We now define the quadrature operators $x_d$ and $x_u$ for the three multipartite states under consideration. Recall that the tilde notation $\tilde{x}_u$ has been introduced earlier to represent sending a mode with quadrature $x_u$ through a channel with transmission $t$, specifically $\tilde{x}_u = \sqrt{t}\,x_u + \sqrt{1-t}\,x_0$, where $x_0$ is the quadrature operator of a vacuum state.

In protocols involving a six-mode graph state, the quadrature of each mode is a linear function composed of quadratures from four squeezed states. Any two modes have two of these squeezed states quadrature in common. Consequently, any two correlated modes of a six-mode graph state can be expressed in terms of the quadrature operators of six squeezed states, all characterized by the same quadrature variance. for example consider the following two quadratures corresponding to deal's mode $x_d$ and mode meant for a remote user $x_u$:
\begin{equation}
x_d = \frac{1}{2}(p^{(2)} + p^{(4)} - x^{(1)} + x^{(3)}),
\end{equation}
\begin{equation}
x_u = \frac{1}{2}(\mp p^{(4)} + p^{(6)} \pm x^{(3)} + x^{(5)}).
\end{equation}

Here, \(x^{(i)}\) denotes a squeezed quadrature with variance \(V\), while \(p^{(i)}\) denotes the corresponding anti-squeezed quadrature with variance \(V_N + \frac{1}{V}\).

Since we are dealing with Gaussian states, straightforward algebraic manipulations yield the following expression for the variance $\mathrm{Var}[x_d \tilde{x}_u]$:
\begin{equation}
\begin{aligned}
\mathrm{Var}[x_d \tilde{x}_u]
&= \tfrac{t}{16} \Bigl(
    5 \bigl(V_N + \tfrac{1}{V}\bigr)^2
    + 5 V^2
    + 6 \bigl(V_N + \tfrac{1}{V}\bigr)V
  \Bigr) \\
&\quad + (1 - t) \left(
    \tfrac{1 + V(V_N + V)}{2V}
  \right).
\end{aligned}
\end{equation}

Turning to the three-mode GHZ state, the relevant quadratures are given by:
\begin{equation}
\begin{aligned}
x_1 &= \frac{\sqrt{2}\,p^{(1)} + x^{(2)} + \sqrt{3}\,x^{(3)}}{\sqrt{6}}, \\
x_2 &= \frac{\sqrt{2}\,x^{(2)} - p^{(1)}}{\sqrt{3}}, \\
x_3 &= \frac{-\sqrt{2}\,p^{(1)} - x^{(2)} + \sqrt{3}\,x^{(3)}}{\sqrt{6}}.
\end{aligned}
\end{equation}

Although two of these modes incorporate quadrature components from three squeezed states each, and one mode has components from only two squeezed states, this difference does not influence the choice of which mode is sent through the transmission channel. The variance \(\mathrm{Var}[x_d \tilde{x}_u]\) remains the same, given by:
\begin{equation}
\begin{aligned}
\mathrm{Var}[x_d \tilde{x}_u]
&= \tfrac{t}{9} \Bigl(
    2\bigl(V_N + \tfrac{1}{V}\bigr)^2
    + 5V^2
    + 2\bigl(V_N + \tfrac{1}{V}\bigr)V
  \Bigr) \\
&\quad + (1 - t)\left(
    \tfrac{1 + V(V_N + 2V)}{3V}
  \right).
\end{aligned}
\end{equation}

For a DAN with a two-mode squeezed vacuum, the relevant quadratures for the dealer and the remote user modes are:
\begin{equation}
x_d = \frac{x^{(1)} + p^{(2)}}{\sqrt{2}}, \quad
x_u = \frac{1}{2}(\sqrt{2}\,p^{(0)} - x^{(1)} + p^{(2)}).
\end{equation}
Similarly, we obtain:
\begin{equation}
\begin{aligned}
\mathrm{Var}[x_d \tilde{x}_u]
&= \tfrac{t}{8}\left(V_N + \tfrac{1}{V}\right)\left(V_N + \tfrac{1}{V} + 1\right) \\
&\quad + V(1+V) \\
&\quad + (1 - t)\left(\tfrac{1 + V(V_N + V)}{2V}\right).
\end{aligned}
\end{equation}

\clearpage

\bibliography{bibliography}

\end{document}